\documentclass[aip,imp,jcp,amsmath,amssymb,superscriptaddress,reprint]{revtex4-1}
\pdfoutput=1

\usepackage{graphicx}
\usepackage{dcolumn}
\usepackage{amsmath}
\usepackage{amsthm}
\usepackage{mathtools}

\newcommand{\lxr}{\langle X \rangle}
\newcommand{\lar}{\langle A \rangle}

\newcommand{\SrA}{S_1}
\newcommand{\SrB}{S_2}

\newcommand{\pax}{\mathcal{P}(A\rightarrow X)}
\newcommand{\pxa}{\mathcal{P}(X\rightarrow A)}

\newcommand{\lP}{\left(}
\newcommand{\rP}{\right)}
\newcommand{\Id}{\textrm{Id}}

\newcommand{\M}{\mathbf{M}}

\newcommand{\bsigma}{\boldsymbol{\sigma}}

\theoremstyle{definition}
\newtheorem{example}{Example}[section]
\newtheorem{rem}{Remark}[section]

\begin{document}

\title{Towards an efficient multiscale modeling of low-dimensional reactive systems: study of numerical closure procedures} 

\author{Giacomo Mazzi}
\email{giacomo.mazzi@cs.kuleuven.be}
\affiliation{Department of Computer Science, KU Leuven, Belgium}
\author{Yannick De Decker}
\affiliation{Interdisciplinary Center for Nonlinear Phenomena and Complex Systems (CENOLI), Universit\'e Libre de Bruxelles, Belgium}
\author{Giovanni Samaey}
\affiliation{Department of Computer Science, KU Leuven, Belgium}



\begin{abstract}
In this paper, we present a study on how to develop an efficient  
multiscale simulation strategy for the dynamics of chemically active
systems on low-dimensional supports. Such reactions are encountered 
in a wide variety of situations, ranging from heterogeneous catalysis 
to electrochemical or (membrane) biological processes, to cite a few. 
We analyzed in this context different techniques within the framework of 
an important multiscale approach known as the equation free method (EFM), which 
``bridges the multiscale gap'' by building 
microscopic configurations using macroscopic-level information only.
We hereby considered two simple reactive processes on a one-dimensional lattice, 
the simplicity of which allowed for an in-depth understanding of the parameters controlling 
the efficiency of this approach. We demonstrate in particular that it is not enough 
to base the EFM on the 
time evolution of the average concentrations of particles on the lattice, but that the time 
evolution of clusters of particles has to be included as well.  
We also show how important it is for the accuracy of this method to carefully choose the procedure 
with which microscopic states are constructed, starting from the measured macroscopic quantities.  
As we also demonstrate that some errors 
cannot be corrected by increasing the number of observed macroscopic variables, 
this work points towards which procedures should be used  in order to generate  
efficient and reliable multiscale simulations of these systems. 
\end{abstract}


\maketitle

\section*{Introduction}


Chemically reacting systems exhibit behaviors that may be described at different levels of modeling. At a fine \emph{microscopic} level, the system consists of a large number of atoms and molecules (particles) of different species, and reactions at this level can be seen as discrete events that occur with some probability when two or more particles are close enough to interact. At a
coarser, \emph{macroscopic} level one only considers the concentration of each species, whose evolution is modeled via a system of ordinary or partial differential equations (depending on whether spatial inhomogeneities are taken into account).
Whereas the microscopic level of description might allow modeling inter-particle interactions in great detail, more macroscopic models are clearly desirable for at least two reasons. First, fine-scale simulation of systems consisting of many particles may be prohibitively expensive, especially when the chemical reactions span a wide range of time-scales. Second, even if feasible, one is usually only interested in the evolution of some coarse-level, macroscopic quantities, such as species concentrations.

When considering only macroscopic quantities, the dynamics of the system is often described using a mean field approximation (MFA), which expresses the law of mass action for the species concentrations; the underlying assumption being that the system is well stirred \cite{nicolisfaraday,provatalattice}.
Unfortunately, in many systems of practical interest, this requirement is not fulfilled.  A typical example is the case of catalytic reactions on surfaces, in which the particles are located on a spatial structure with low dimensionality, such as a lattice. Then, the location of the potentially reacting particles becomes important, as chemical reactions can only fire if the reactants are physically close to each other and mixing is so inefficient that local fluctuations of composition are not easily wiped out. 
A possible approach is to expand the set of macroscopic state variables, taking into account particle correlations. As for the mean field equations, closed equations for the pair correlations can only be obtained using additional assumptions. Several analytical closure approximations have been developed such as the quasi-chemical approximation \cite{quasichem}; additional references are given in Section~\ref{sec:macro-vars}. A drawback of analytic closure approximations is that they may depend strongly on the system considered and when, besides pair correlations, other additional variables are needed, the equations quickly become intractable.

When no closed macroscopic model with sufficient accuracy is available, one needs to resort to an individual-based stochastic particle simulation. In this case, the microscopic state of the system is updated via a kinetic Monte Carlo (kMC) algorithm according to a given set of reaction laws, such as the stochastic simulation algorithm (SSA) \cite{SSA}, with adaptations to also take into account the underlying lattice structure in the aforementioned case of surface reactions.
These simulations are often time consuming and  several methods have been developed that take advantage of disparate reaction rates to accelerate them. We mention, for instance, tau-leaping \cite{tauleaping}, R-leaping \cite{r-leaping}, and nested SSA \cite{nestedSSA}. However, most of these methods are not easily implemented in lattice kMC, unless other approximations are also taken into account; see  [\onlinecite{tauleaplattice}] for an example which makes use of a type of coarse-graining. For a general review of kinetic Monte Carlo methods in this context, see [\onlinecite{chatterjeeoverview}].

There exists, however, an alternative way of extracting the macroscopic dynamics. Computational strategies have recently been developed that are capable of accelerating individual-based particle simulations by introducing on-the-fly \emph{numerical} approximate macroscopic closures: we mention equation-free \cite{efm2003,annrevEFM} and heterogeneous multiscale methods (HMM) \cite{EEng03,EEng07}. In both approaches, the crucial steps are (i) to decide on the set of macroscopic state variables that are sufficient to determine future evolution of the system, and (ii) to define an operator that generates a microscopic state corresponding to a given macroscopic state. This second step is actually equivalent to prescribing the closure approximation; in the equation-free, resp.~HMM, frameworks, this step is called lifting, resp.~reconstruction. The converse operator (mapping a microscopic state to a macroscopic one) is called restriction in both frameworks. Once these operators are available, one can construct a \emph{coarse time-stepper}, which evolves the chosen macroscopic state variables over some time interval by means of a three-step procedure: (i) lifting; (ii) microscopic simulation; and (iii) restriction.

Equation-free methods have already been used to perform  numerical bifurcation analyses of lattice chemical systems
 \cite{makeev,makeev2009}. In [\onlinecite{makeev}], one studies a CO oxidation process on a lattice for which a 
bistability observed in simulations is not fully reproduced by a mean-field or a quasi-chemical approximation.  
 Using a coarse time-stepper in which the macroscopic state is given by 
the concentrations of all species, however, the observed macroscopic bifurcation results can be completely reproduced 
and analyzed. This result is the consequence of two fundamental underlying properties of the system: (i) the coverages 
by themselves are indeed sufficient to predict macroscopic evolution of the system; (ii) given the global concentrations, one can construct 
a lifting operator that generates appropriate initial conditions for individual lattice sites. In [\onlinecite{makeev2009}] NO reduction
 by CO over a $Pt(100)$ surface is also analyzed with the same approach.  In both cases, the lifting 
equilibrates the lattice by exploiting a separation in time scales between (fast) diffusion on the lattice and (slower) chemical 
reactions. As a result, all high order spatial correlations are slaved to the concentrations.  This separation of time scales allows 
for what the authors call a property of ``healing": incorrectly initialized higher order correlations quickly relax to a correct 
functional of the coverages.  

The present paper assesses the efficiency of lifting/reconstruction techniques in the case of chemical reactions on lattices; the procedure we propose, however, could be applied to many multiscale models, see e.g. [\onlinecite{samaey2011}] for a similar study in the context of polymeric fluids. In particular, we investigate systems that are subject to a set of chemical reactions in a crowded environment that makes diffusion ineffective. Such situation puts in danger the aforementioned diffusion-induced separation of time scales.


As already pointed out in [\onlinecite{makeev}], two problems are expected to arise in this context. First, we will more than probably need to augment the set of macroscopic state variables, as an approximate closure on the level of coverages alone will generally not suffice.  Second, in the absence of a clear time-scale separation, artifacts introduced during the lifting step cannot be expected to ``heal'' completely. Additionally, we are not only interested in capturing the correct steady state behavior in the present paper, but also in recovering the (transient) macroscopic dynamics. 
In this view, the main contribution of the present paper is twofold:
\begin{itemize}
\item  From an algorithmic point of view, we develop a numerical closure strategy to recover the macroscopic (transient) dynamics of two specific chemical systems with nonlinear reactions on a one-dimensional lattice, defined in Section~\ref{sec:mod_prob}.  We introduce a hierarchy of macroscopic state variables for which the implementation of a consistent lifting operator is feasible (Section~\ref{sec:ef}) and consider several options to initialize the remaining degrees of freedom (Section~\ref{sec:lift_proc}).  
\item From a computational physics point of view, we analyze the effect of the different lifting strategies (corresponding to different closure approximations) on the macroscopic evolution (Section~\ref{sec:num_exper}). This study reveals that both the number and type of macroscopic state variables and the choice of the lifting operator (for given macroscopic state variables) can have a significant effect on the observed macroscopic evolution and hence, on the accuracy of the multiscale approach. 
\end{itemize}
We analyzed this on the case of two chemical systems with nonlinear reactions on a one--dimensional lattice. Our choice of one-dimensional systems is justified not only by a corresponding simple analysis, but also because this setting has the additional advantage that effects due to the low dimensionality are more pronounced. Consequently, if a procedure provides accurate results in one space dimension, we expect it to also perform well in higher dimensions. 

The paper is organized as follows: in Section \ref{sec:mod_prob}, we present the model problems considered in this work. We describe how these systems are modeled, as well as how they are simulated numerically.
In Section \ref{sec:ef}, we introduce the multiscale framework, including the hierarchy of macroscopic state variables (Section~\ref{sec:macro-vars}), the different lifting and restriction operators (Section~\ref{sec:lift-restrict}), and the numerical closure algorithm (Section~\ref{sec:algo}).
In Section \ref{sec:lift_proc} we describe the different lifting procedures.
The numerical results we have obtained on two model problems are presented in Section\ref{sec:num_exper}.  We conclude in Section~\ref{sec:concl}, where we also outline open questions and directions for future research.

\section{Model problems and coarse-scale approximations}\label{sec:mod_prob}

\subsection{Chemical reactions on one-dimensional lattices}

We consider a one-dimensional spatial domain with $N$ lattice sites and periodic boundary conditions (i.e., a ring).  Each lattice site is occupied by exactly one particle.  
For simplicity, we consider only two particle species $X$ and $A$.
The dynamics of the system is governed by a set of $I$ reactions, with associated reaction rates $k_i$ ($1\le i \le I$). Reactions can only occur if the reacting particles are in adjacent sites.
We adopt a slight modification of the notations that were introduced previously in the literature\cite{glauber,tretprovata,prakschlogl}; we define the microscopic state at time $t$ as $\boldsymbol{\sigma}(t)=\{\sigma_1(t),\sigma_2(t),\ldots,\sigma_N(t)\}$ where each variable $\sigma_n(t)=\pm 1$, depending on whether the lattice site $n$ is occupied by $A$ ($\sigma_n(t)=+1$) or $X$ ($\sigma_n(t)=-1$). 
We also define by
\begin{equation}\label{eq:coverage}
a(t):=\frac{1}{2}\left( \frac{1}{N}\sum_{n=1}^N \sigma_n(t)+1\right),
\end{equation}
the coverage of species $A$, and correspondingly $x=1-a$ the coverage of species $X$.

The system state evolves in time as a consequence of chemical reactions. Throughout this text, we will consider two particular example systems:
\begin{example}[Schl\"ogl model]\label{ex:schl}
	Consider the following two-species system, in which two reversible reactions are possible,
	 \begin{equation}\label{kin_law_schlogl}
	\begin{array}{ccc}
	XAX &\xrightleftharpoons[k_2]{k_1} &XXX,\\
	X&\xrightleftharpoons[k_4]{k_3}&A,\\
	\end{array}
	\end{equation}
	where the notation in the first line is chosen to emphasize that only elements surrounded by two $X$ may react. 
This type of system is a two-species restriction of the popular Schl\"{o}gl model (usually called Schl\"{o}gl's second model \cite{prakschlogl}), and it is widely used as prototype for study of non-equilibrium systems on constrained configurations, such as catalytic reactions on surfaces \cite{makeev2002,dedecker2010}.  Similar equations have been proposed to describe disease spreading in epidemic models \cite{Sirsepidemic}.
\end{example}

\begin{example}[Simple trimolecular reaction]\label{ex:trim}
 We may simplify the above prototypical example even further by only retaining the trimolecular reaction, i.e.,
 \begin{equation}\label{kin_law2}
XAX \xrightleftharpoons[k_2]{k_1} XXX,
\end{equation}
where we assume $k_1=k_2=1$. This model contains the simplest mechanism to obtain a macroscopic behavior that non-trivially depends on the local configurations\cite{provataturner,tretprovata}.  Note that this example arises when considering Schl\"ogl's second model in the limit where $k_3=k_4=0$. 
\end{example}

\subsection{Stochastic simulation}

A common way to simulate chemical reactions on a lattice is via stochastic simulation, in which one computes one realization of the stochastic process as a function of time.  As in most of the earlier literature for Example~\ref{ex:schl} and Example~\ref{ex:trim} \cite{tretprovata,provataturner}, we apply a kinetic Monte Carlo (kMC) method with an accept-reject strategy (a so-called null-event kMC, see, e.g., \cite{nulleventKMC,chatterjeeoverview}). As for the classical stochastic simulation algorithm (SSA) \cite{SSA} for well-stirred systems (which is rejection-free), some modifications are required, as the reaction probabilities not only depend on the selected reaction but also on the chosen lattice site. 

Any kMC method depends on the definition of a (possibly time-dependent) propensity function $\Gamma_i(t)$ per reaction ($1 \le i \le I$); the quantity $\Gamma_i(t) dt$ represents the probability of reaction $i$ occurring in the time interval $[t,t+dt]$. The main modification to accommodate for the simulation of reactions on low-dimensional lattices is that the propensity functions, which are spatially independent for well-stirred systems, now also depend on the reaction site $n=1,\ldots,N$. Therefore, at each moment in time, we have $N\times I$ propensity functions, which we denote as $\Gamma_{i,n}$, with $1\le n \le N$. Moreover, the propensity functions do not depend on the total coverages of each of the species, but on the states of the individual lattice sites. For Example~\ref{ex:schl}, $\Gamma_{i,n}$ may be derived from (\ref{kin_law_schlogl}) as
\begin{small}
\begin{align}\label{gamma}
XAX \xrightharpoonup{k_1} XXX,\quad \Gamma_{1,n}(t)&=\frac{1+\sigma_n(t)}{2}k_1\frac{(1-\sigma_{n-1}(t))}{2}\frac{(1-\sigma_{n+1}(t))}{2},\nonumber\\
XXX \xrightharpoonup{k_2} XAX,\quad \Gamma_{2,n}(t)&=\frac{1-\sigma_n(t)}{2}k_2\frac{(1-\sigma_{n-1}(t))}{2}\frac{(1-\sigma_{n+1}(t))}{2},\nonumber\\
X\xrightharpoonup{k_3}A,\quad \Gamma_{3,n}(t)&=\frac{1-\sigma_n(t)}{2}k_3,\\
A\xrightharpoonup{k_4}X,\quad\Gamma_{4,n}(t)&=\frac{1+\sigma_n(t)}{2}k_4.\nonumber 
\end{align}
\end{small}
We further introduce the total propensity for a lattice site $n$ as the sum of all reaction propensities for that lattice site, $$\Gamma_{\textrm{tot},n}(t)=\sum_{i=1}^I\Gamma_{i,n}(t),$$ where  $\Gamma_{\textrm{tot},n}(t)dt$ represents the probability that \emph{any} reaction will occur at the site $n$ in the time interval $[t,t+dt]$.
The maximal total propensity over all lattice sites is defined as $\Gamma_{\max}(t)=\max_n \Gamma_{\textrm{tot},n}(t)$. 
Finally, we will also use the total propensity of a reaction $i$, which is defined as the sum of the reaction propensities for that reaction over all lattice sites $$\Gamma_{i,\textrm{tot}}(t)=\sum_{n=1}^N\Gamma_{i,n}(t).$$
One time step of the algorithm then proceeds as follows \cite{chatterjeeoverview}:
\begin{enumerate}	
\item A lattice site $m$ is randomly chosen (from a discrete uniform distribution on $\left\{1,\ldots,N\right\}$), and all propensity functions $\Gamma_{i,m}(t)$ for the reactions are calculated at that site, and renormalized to obtain $\tilde{\Gamma}_{i,m}(t)=\Gamma_{i,m}(t)/\Gamma_{\max}(t)$.  Note that, as a consequence of the renormalization, we have the property  
\begin{equation}
\sum_{i=1}^{I}\tilde{\Gamma}_{i,m}\leq 1.
\end{equation}
\item Next, a reaction is randomly chosen according to the propensity functions. First generate a random number $\xi$ from the uniform distribution on $[0,1]$. The chosen reaction is then given as $j$ such that
\begin{equation}\label{step2ssa}
\sum_{i=1}^{j-1} \tilde{\Gamma}_{i,m}(t) < \xi \leq\sum_{i=1}^{j}\tilde{\Gamma}_{i,m}(t).
\end{equation}
When $\Gamma_{\textrm{tot},m}(t)<1$, we might have that $\Gamma_{\textrm{tot},m}(t)/\Gamma_{\max}(t)<\xi \le 1$. In that case, there is no reaction $j\le I$ for which the above equation is satisfied; this situation corresponds to a null-event, and no reaction is performed. By convention, we set $j=I+1$ in this case.
\item Compute the time increment
$$
\delta t_{j}(t)=\begin{cases}
1/\Gamma_{j,\textrm{tot}}(t),& 1\le j \le I \\
0 & j = I+1
\end{cases}$$
and advance the system clock, $t=t+\delta t_j(t)$.
\end{enumerate}
The last step differs from SSA, as there the size of $\delta t_j$ is evaluated using another random generated number. 

\begin{rem}In the case of Example~\ref{ex:trim}, this algorithm may be simplified considerably. We can then pick randomly a lattice site $m$ (again according to a discrete uniform distribution). If $\sigma_{m-1}(t)=\sigma_{m+1}(t)=-1$, then we change the state of that site $\sigma_m(t+\delta t)=-\sigma_m(t)$. For this example, $\delta t$ is constant as a function of time, and independent of the chosen reaction.
\end{rem}

\subsection{Master equation and approximation of macroscopic dynamics}

Instead of performing a \emph{stochastic} simulation that governs the evolution the configuration for an individual realization of the system, one can also describe the \emph{deterministic} evolution of the \emph{probability distribution} of the realizations over the configuration space, which is governed by the master equation.  While the computational cost of this approach is prohibitive for practical purposes, this description is useful to derive approximate macroscopic models.

Let us introduce the probability distribution $\mathcal{P}(\boldsymbol{\sigma},t)$  \cite{glauber,nicolisfaraday}, and corresponding master equation
\begin{equation}\label{master}
\frac{d \mathcal{P}(\boldsymbol{\sigma},t)}{dt}=\sum_{n=1}^N\left[ w_n(\boldsymbol{\bar{\sigma}}^n)\mathcal{P}(\boldsymbol{\bar{\sigma}}^n,t)- w_n(\boldsymbol{\sigma})\mathcal{P}(\boldsymbol{\sigma},t)\right],
\end{equation} 
in which $\boldsymbol{\sigma}=\{\sigma_1,\ldots,\sigma_n,\ldots,\sigma_N\}$, and each $\boldsymbol{\bar{\sigma}}^n=\{\sigma_1,\ldots,\sigma_{n-1},-\sigma_n,\sigma_{n+1},\ldots,\sigma_N\}$ differs from $\boldsymbol{\sigma}$ in exactly one lattice site. The rates $w_n(\boldsymbol{\sigma})$ are the probability per unit time that the $n$-th particle flips from $\sigma_n$ to $-\sigma_n$, given that the current state of the whole system is $\boldsymbol{\sigma}$.  For the systems we consider in the present paper, the form of $w_n$ depends on the different reaction laws and on the nearest neighborhood $\left\{\sigma_{n-1},\sigma_{n+1}\right\}$.  For instance, given the reaction laws of Example~\ref{ex:schl}, we get the reaction probabilities per site\cite{prakschloglA}, 
\begin{small}
\begin{equation}\label{w_jex}
\begin{aligned}
w_n\left(\boldsymbol{\sigma}\right)&=\left[\frac{k_1+k_2}{2}+\frac{k_1-k_2}{2}\sigma_n\right]\frac{1-\sigma_{n-1}}{2}\frac{1-\sigma_{n+1}}{2} \\&
+\frac{k_4+k_3}{2}+\frac{k_4-k_3}{2}\sigma_n.
\end{aligned}
\end{equation} 
\end{small}
The $w_n$ are also related with the propensity functions (\ref{gamma}) as $w_n(\boldsymbol{\sigma(t)})=\Gamma_{{\rm tot},n}(t)$.
Note that the first two terms, which refer to the first reaction, depend on the state of the nearest neighbors, in contrast to the last two terms, which refer to the second reaction. 

For a majority of systems,  the master equation~\eqref{master} is too complicated to be solved directly. However, when introducing suitable approximations, it is possible to extract from it equations of motions for appropriately chosen macroscopic quantities (such as coverages).
The simplest approximate macroscopic model is the so-called mean field approximation, where, by substituting all the high order correlations with an average ``mean field'' interaction, one writes a set of ordinary differential equations for the coverages \cite{prakschlogl,prakschloglA} that can be simulated with drastically reduced computational complexity.

Assuming translational invariance, 
it was shown in \cite{prakschlogl,prakschloglA} that the mean-field evolution equation for $q(t)=\left< \sigma_{n} (t)\right>$
is here given by
\begin{align}\label{mfsch}
\frac{dq}{dt}&=-\frac{1}{4}(k_1-k_2)+(k_3-k_4)+\left(\frac{1}{4}(k_1-3k_2)+k_3+k_4\right)q\nonumber \\
&+\frac{1}{4}(k_1+3k_2)q^2-\frac{1}{4}(k_1+k_2)q^3,
\end{align}  
which may be written in terms of one of the coverages (here $x=(1-q)/2$), and reads
\begin{equation}
\frac{dx}{dt}=-(k_1+k_2)x^3+k_1 x^2-(k_3+k_4)x+k_4.
\end{equation}
If we set $k_3=k_4=0$, we get the equations for Example \ref{ex:trim},
\begin{equation}\label{MF}
\frac{d x}{dt}=-k_1 x^3 + k_2 a x^2.
\end{equation}

The mean-field approximation is based on neglecting of all pair (and higher order) correlations. It is therefore expected to perform poorly in the systems of interest in this paper: in fact, this level of description even yields  equilibrium values that differ substantially from those obtained by the microscopic dynamics (\ref{kin_law2}) \cite{provatalattice}. It nevertheless represents a suitable starting point for more involved approximation strategies.  For instance, it is possible to also consider pair correlations via the Ursell expansion \cite{abad}. This approach consists in a development of pair, triplet, quadruplet, etc. correlation functions from which one can systematically extract finer and finer descriptions as higher order terms are kept.  Another approach consists in expressing high order moments directly as functional of the lower ones. 
This is the case for the popular Kirkwood approximation scheme \cite{evans} or the (less well-known) Bethe-type ansatz \cite{prakschlogl}. The efficiency of several of these approaches for the models under consideration here are discussed in \onlinecite{prakschlogl} where it was shown that 3-points correlations play  as an important role as pair correlations in the dynamics. In any case, despite the simplicity of the considered model, the derivation and analysis of higher order approximations already revealed to be quite tedious and intricate. This justifies the development of numerical closure techniques, which we will discuss now.

\section{Numerical closure algorithm}\label{sec:ef}

We now describe the numerical closure algorithm that is the focus of the present paper. We first define a hierarchy of macroscopic state variables (Section~\ref{sec:macro-vars}). Subsequently, we introduce numerical \emph{lifting}  and \emph{restriction} procedures to map macroscopic state variables to microscopic realizations, and vice versa (Section~\ref{sec:lift-restrict}). Finally, we propose a \emph{coarse time-stepper} that mimics the evolution of the corresponding unavailable macroscopic model \cite{efm2003,annrevEFM}. 

\subsection{A hierarchy of macroscopic state variables}\label{sec:macro-vars}

As in the mean-field approximation, we first introduce a macroscopic state variable $a(t)$, see equation~(\ref{eq:coverage}), that represents the coverage of species $A$; recall that the coverage of species $X$ is obtained as $x=1-a$. In addition to the coverage, we introduce a hierarchy of macroscopic state variables. Each macroscopic state variable $M^l_{A}$, $1\le l \le L_A$, (respectively $M^l_X$, $1\le l \le L_X$) is defined as the number of clusters of particles $A$ (respectively $X$) of \emph{exactly} length $l$ up to a maximum length of $L_A$ (respectively $L_X$). Explicitly, these additional state variables can be written as
\begin{equation}\label{mxjmaj}\begin{aligned}
M_A^l&=\frac{1}{2^{l+2}}\sum_{n=1}^N (\sigma_{n-1}-1)(\sigma_{n+l}-1)\prod_{k=0}^{l-1}(\sigma_{n+k}+1),\\
M_X^l&=\frac{1}{2^{l+2}}\sum_{n=1}^N (\sigma_{n-1}+1)(\sigma_{n+l}+1)\prod_{k=0}^{l-1}(\sigma_{n+k}-1),
\end{aligned}
\end{equation}
where periodic boundary condition on the index $n$ are assumed.  The complete set of macroscopic state variables is then denoted as  $$\M_{L_A,L_X}=\{a,M^1_A,\ldots,M_A^{L_A},M_X^{1},\ldots,M_X^{L_X}\}.$$

\begin{rem} We stress that the above definition is slightly different from other definitions of macroscopic state variables based on cluster counting already presented in the literature, the most notable being the Kirkwood approximation \cite{evans}. For instance, the variable $M^1_A$ only takes into account \emph{single} A's (surrounded on both sides by $X$). Also, a cluster of length $L_{\{A,X\}}=3$ is not counted simultaneously as a two clusters of length $2$, as is done in the Kirkwood approximation;  we only count structures of the form $AXXA$ to determine the number of clusters of length $2$.  To avoid potential confusion, we introduce the notation
	\begin{equation}\label{eq:kirkwood}
		\langle XXX \rangle_{\boldsymbol{\sigma}} = \frac{1}{2^{l+2}}\sum_{n=1}^N \prod_{k=0}^{l-1}(\sigma_{n+k}-1),
	\end{equation}
	to introduce the number of lattice sites occupied by species $X$, and also surrounded by species $X$, as it would appear in the Kirkwood approximation. 
\end{rem}

As will become clear in Section~\ref{sec:lift-restrict}, this hierarchy of macroscopic state variables greatly simplifies the implementation of a consistent lifting strategy, without introducing additional complexity into the restriction.

\subsection{Lifting and restriction}\label{sec:lift-restrict}

We introduce two operators that make the transition between microscopic and macroscopic state variables.
We define a \emph{lifting operator},
\begin{equation}\label{eq:intro_lifting}
\mathcal{L}: \M_{L_A,L_X} \mapsto \boldsymbol{\sigma},
\end{equation}
which maps a macroscopic state to a realization of a chain of length $N$, and an associated
\emph{restriction operator},
\begin{equation}\label{eq:intro_restriction}
\mathcal{R}: \boldsymbol{\sigma} \mapsto \M_{L_A,L_X},
\end{equation}
which maps a microscopic realization to the corresponding macroscopic state.
The restriction operator is readily defined using formulas \eqref{mxjmaj}.
The lifting step, however, is more involved. Besides the obvious condition that the lifting needs to be consistent, i.e.,  
$$\mathcal{R} \circ \mathcal{L} \equiv \Id,$$ one also needs to ensure that microscopic configurations that result from lifting do not contain undesirable artifacts, and actually represent realistic microscopic configurations. For such purpose we introduce the concept of dynamical equivalence to compare different microscopic configurations.

For the reaction laws (\ref{kin_law_schlogl}), we say that two microscopic configurations $\boldsymbol{\sigma}_1$ and $\boldsymbol{\sigma}_2$ are equivalent if they have the same number of possibly reacting lattice sites for  each of the possible reactions, i.e., if
\begin{align}
\langle XXX\rangle_{\boldsymbol{\sigma}_1}&=\langle XXX\rangle_{\boldsymbol{\sigma}_2},\nonumber \\ 
M_A^1(\boldsymbol{\sigma}_1)&= M_A^1(\boldsymbol{\sigma}_2), \label{eq:equiv}\\
a(\boldsymbol{\sigma}_1)&=a(\boldsymbol{\sigma}_1),\nonumber
\end{align}
where the notation in the first equality needs to be understood in the Kirkwood sense, see equation~\eqref{eq:kirkwood}.
The last two conditions are satisfied for microscopic configurations that have the same macroscopic state, for any $L_X$ and $L_A \ge 1$, while the first condition is satisfied exactly only by conserving all the clusters (for instance, by choosing $L_X=L_A=N$). In this case, lifting corresponds to a random shuffle of clusters: the number of cluster is conserved, but not their exact locations. So, using the definition of equivalence introduced above, one has
\begin{equation}\label{fullexp}
\bsigma\equiv\mathcal{L}(\M_{N,N}(\boldsymbol{\sigma})).
\end{equation}

When decreasing the number of macroscopic state variables (by choosing smaller $L_X$ and $L_A$), the numbers of longer clusters will not be preserved during lifting. Instead, we only conserve the \emph{total number} of elements of species $X$ and species $A$ that are not contained within the counted clusters. These remainders can be computed as
\begin{equation}\label{eq:remainder}
	\begin{aligned}
r_A=&\;a \, N -\sum_{l=1}^{L_A} l M_A^l,\\
r_X=&\left(N-a\,N\right)-\sum_{l=1}^{L_X} l M_X^l.
\end{aligned}
\end{equation}
During a lifting procedure, one will typically choose a limited amount of macroscopic variables from the available set to construct the microstates. This reconstruction can be done in many different ways, depending on how one treats the above-defined remainders to create a lattice state that is compatible with the macroscopic constraints. We here emphasize that each such lifting policy actually corresponds to an implicit choice of closure. As  it can clearly be seen now, lifting indeed corresponds to reconstructing higher order moments, starting from low order ones.
The investigation of different choices for the lifting is exactly the focus of the present paper. In particular, the main goal is to investigate which lifting operator recovers the macroscopic dynamics of the system with a minimal number of macroscopic variables. In Section~\ref{sec:lift_proc}, a number of choices will be presented. Numerical results using the resulting lifting operators are given in Section~\ref{sec:num_exper}. 
 

\subsection{The numerical closure algorithm}\label{sec:algo}

We are now ready to formulate the numerical closure algorithm that will be used to simulate time evolution of the macroscopic state variables. Given an initial condition for the macroscopic state variables $\M(t^*)$ at
time $t^*$, we define \emph{coarse time-stepper} as the following three-step procedure: 
\begin{itemize}
\item[(i)]	\emph{Lifting}, {\em i.e.}\, the
creation of initial conditions $$\bsigma(t^*)=\mathcal{L}(\M_{L_A,L_X}(t^*)),$$ for the microscopic model, consistent with the macroscopic state  $\M_{L_A,L_X}$ at time $t^*$.
\item[(ii)] \emph{Simulation} of the evolution of the microscopic state over a time interval $[t^*,t^*+ \delta t]$.
\item[(iii)] \emph{Restriction}, {\em i.e.}\, the observation (estimation) of the macroscopic state
at $t^*+ \delta t$:
$$\M_{L_A,L_X}(t^*+  \delta t)=\mathcal{R}(\bsigma(t^*+ \delta t)).$$
\end{itemize}
The coarse time-stepper can be used to accelerate simulation in time using coarse projective integration algorithms \cite{annrevEFM}, or to perform a numerical bifurcation analysis \cite{efm2003}.
In this paper, we will use the coarse time-stepper to study how different choices for the numbers $L_A$ and $L_X$ of macroscopic state variable $\M_{L_A,L_X}$, and of the specifics of the lifting (determining the distribution of the remainders $r_A$ and $r_X$) introduce a closure approximation (i.e., affect the macroscopic dynamics).
To this end, we compare the time series of the macroscopic state variables $\M_{L_A,L_X}$ obtained via the numerical closure algorithm with those obtained by a complete microscopic simulation, $\M_{L_A,L_X}(t)=\mathcal{R}(\sigma(t))$. A similar study has been performed for a model for polymeric fluids \cite{samaey2011}.

\section{Lifting procedures}\label{sec:lift_proc}

Let us now detail the specific lifting operators that we propose in this paper. Each lifting operator consists of two steps: (i) positioning of the conserved clusters of species $X$ and $A$, as specified by the macroscopic state variables $\M_{L_A,L_X}$; and (ii) distribution of the remaining $r_A$ (respectively $r_X$) elements of species $A$ (respectively $X$). 

The first step is common to all developed strategies, limiting differences in lifting operators to the distribution of the remainders.  Each macroscopic state variable $M^l_A$ ($1 \le l \le L_A$), respectively $M^l_X$ ($1 \le l \le L_X$), refers to the number of clusters of length $l$ for species $A$, respectively $X$. We gather these clusters in two sets: $B_A= \{ A,\ldots,A,AA,\ldots AA, AAA,\ldots, AAA,\ldots\}$, and similarly $B_X$, 
and then select elements from $B_A$ and $B_X$ alternatingly until one of the two sets is empty.  We further introduce the notation 
\begin{equation}
m_A=\left|B_A\right|= \sum_{l=1}^{L_A} M^l_A ,\qquad m_X=\left|B_X\right|=\sum_{l=1}^{L_X} M^l_X,
\end{equation}
for the cardinality of both sets.  
Since usually $m_A\neq m_X$,  some elements from $r_A$ or $r_X$ will need to be used to ensure that we can place all clusters that are accounted for in the macroscopic state variables (see below). In the remainder, we assume that $m_X>m_A$, the argument for the alternative case is identical.

Given the definition of the macroscopic state variables,  we need to group elements contained in the remainders $r_A$ (respectively $r_X$) in blocks longer than $L_A$ (respectively $L_X$). The lifting procedures described below differ exactly in the way these remaining elements are taken in account.
Let us define by $C_A$ (respectively $C_X$) the number of blocks containing only elements of species  $A$ ($X$ respectively) which will be created with the $r_A$ (respectively $r_X$) remaining elements.  We propose the following alternatives for the choice of $C_A$ and $C_X$:
\begin{enumerate}
\item {\bf lift$_A$} Given $r_A$ and $r_X$, we build as many blocks as possible, i.e., $$C_X=\left\lfloor\dfrac{r_X}{L_X+1}\right\rfloor \text{ and } C_A=\left\lfloor\frac{r_A}{L_A+1}\right\rfloor.$$ Then, most of the $C_A$ blocks of species $A$ will have length $L_A+1$; others will be longer because of the remainder of the division between $r_A$ and $L_A+1$. The same is true for the clusters of species $X$. Once these blocks have been created we have only to decide how to place them on the lattice. Having most of the blocks of equal length $L_X+1$ may introduce artifacts on the distribution of clusters resulting from the lifting. To partially increase the randomness of such distribution we  modify the equations for $C_X$ and $C_A$ by taking them to a random integer number picked by a uniform distribution between 
\begin{equation}\label{liftACACX}
\left(\frac{r_X}{L_X+2},\frac{r_X}{L_X+1}\right) \text{ and }\left(\frac{r_A}{L_A+2},\frac{r_A}{L_A+1}\right)
\end{equation}
In such way the length of the blocks is distributed more randomly. We show in the next section how this choice for $C_X$ and $C_A$ still introduces artificial oscillations in the distribution of clusters.

\item  {\bf lift$_B$} Instead of building as many blocks as possible, we might generate a more realistic number of them. To this end, we derive an upper and a lower limit for $C_A$ and $C_X$, and we additionally use the constraint that the total number of clusters for both species is equal, i.e., 
\begin{equation}\label{eq:nb_clust_B}
    C_A+m_A=C_X+m_X.
\end{equation} Then, the maximum number of clusters for species $X$ can be derived to be
\begin{equation}\label{maxC}
\begin{aligned}
C_X^{\rm max}&=\min\Big(\left\lfloor\frac{r_X}{L_X+1}\right\rfloor, \left\lfloor\frac{r_A}{L_A+1}\right\rfloor  \\&
+\max(m_A-m_X,0) -\max(m_X-m_A,0)\Big), \nonumber
\end{aligned}
\end{equation}
whereas for the minimum we have
\begin{equation}\label{minC}
C_X^{\min}=\max(\delta_{r{_X}}, m_A-m_X),
\end{equation}
where $\delta_{r{_X}}=1$ if $r_X>0$ and $0$ otherwise; $\delta_{r_X}$ takes into account the unlikely event where all the elements $X$ are contained into $B_X$. Similar formulas can be derived for $C_A^{\max}$ and $C_A^{\min}$.  
We then pick a number of clusters $C_X$ randomly, based on a uniform distribution between the corresponding minimum and maximum, and infer the number of clusters $C_A$ from \eqref{eq:nb_clust_B}.

\item {\bf lift$_C$} In \textbf{lift}$_B$, the lower limits $C^{\min}_{A,X}$ can be very conservative, in fact this limit refers to a configuration where all the elements of $r_X$ (respectively $r_A$) are packed together in a single long cluster, such that the resulting lifting operator artificially creates a low number of clusters.  This corresponds to clusters which are artificially long. We therefore also investigate a lifting in which the minimum number of clusters is based on the derivation of a ``reasonable'' maximal cluster length, based on the specific values of $r_X$ and $r_A$, from which the minimal number of clusters $C^{\min}_{A,X}$ then is derived,  the argument  based on the probabilistic reasoning.
In the next section, we write an explicit estimate for the example \ref{ex:schl}.
\end{enumerate}

Once the numbers of clusters have been determined, and the corresponding individual clusters have been created, they need to be positioned on the lattice. Let us denote by $B_A^r$ (respectively $B_X^r$) the sets containing the clusters built out of the remainders of each species. Recall that, given the assumption that $m_X>m_A$, we still need to place, besides the clusters corresponding to the remainders, $m_X-m_A$ clusters of species $X$ corresponding to the macroscopic state variables.  To this end, we continue as before, only now we alternatingly pick an element from $B_X$ and $B_A^r$.  After that step, for {\bf lift}$_A$ the remaining $n^r:=C_A+C_X-(m_X-m_A)$ clusters in $B_A^r\cup B_X^r$ are randomly picked and appended to the configuration; for {\bf lift}$_B$ and {\bf lift}$_C$ instead we have an equal number of elements remaining in $B_A^r$ and $B_X^r$ so it is sufficient to pick in the alternate way from the two sets.

Note that from the cluster definition \eqref{mxjmaj} when reconstructing we impose adjacent clusters to superimpose at the head and the tail. 
The case where $m_X>m_A$ can be handled identically.
\begin{rem}[Lifting based on coverage]\label{rem:lift_r} When choosing $L_A=L_X=0$, the only macroscopic state variable is the coverage $a$, i.e. $\M=\left\{a\right\}$. In that case, the lifting generates a configuration $\bsigma$ in which the position of the $N\times a$ elements of species $A$ is completely random.  This method is labeled as \textbf{lift}$_R$.
\end{rem}

\section{Numerical experiments}\label{sec:num_exper}

In this section, we illustrate the use and the efficiency of the different lifting operators defined above. We apply them on the two examples 
introduced in Section \ref{sec:mod_prob}, analyze and quantify the deviations from the original microscopic simulations and discuss
the reasons behind the observed discrepancies. 

\subsection{Example \ref{ex:schl}}\label{sec:num_schl}

\begin{figure}
 \begin{centering}
\includegraphics[width=0.49\textwidth]{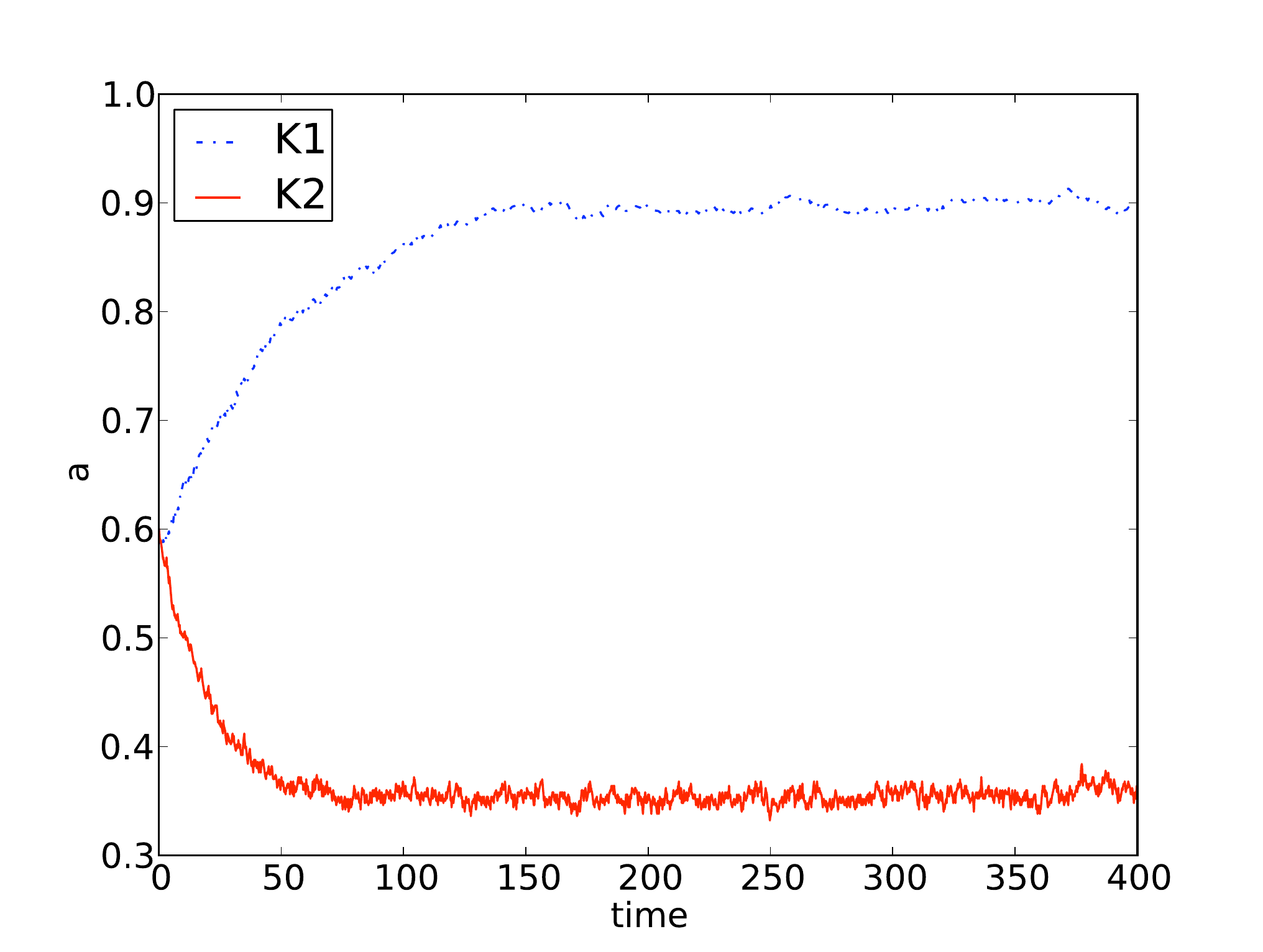}  
\caption{\label{fig:only_micro}Results for the microscopic dynamics for $K=K_1$ and $K=K_2$ reaction rates.}
 \end{centering}
 \end{figure}

We study the system \eqref{kin_law_schlogl} with two different reaction rates sets: $K_1=[k_1=2,k_2=1,k_3=0.1,k_4=0.01]$ and $K_2=[k_1=1,k_2=2,k_3=0.01,k_4=0.1]$.  In both cases, the single-element reactions are slower than the reactions including nearest-neighbor interactions. The choice of these specific parameter sets is motivated by the differences in the resulting equilibrium values. For the reaction rates $K_1$,  the concentration at equilibrium is $a_{\rm eq} \simeq 0.9$; for the reaction rates $K_2$, is $a_{\rm eq} \simeq 0.35$, see Figure~\ref{fig:only_micro}. Unless otherwise stated, all simulations are performed on a lattice with $N=2000$ sites for a time interval $T = [0,400]$. We show in the next sections how the different equilibrium values effect the quality of the various liftings. We first assess the accuracy of the lifting procedure \textbf{lift}$_R$ whose macroscopic description is solely based on coverages (Section~\ref{sec:lift_r}). We then introduce the augmented macroscopic variables set (Section \ref{sec:macro_state}) and  perform a numerical test using each of the discussed lifting strategies (Sections \ref{sec:lift_A}--\ref{sec:lift_c}).  The test consists of simulating using the coarse time-stepper using the minimal possible time step. Hence, after every microscopic time step, we first restrict to the selected macroscopic state variables, after which we lift back to a microscopic configuration. As a result, the observed macroscopic time evolution will mimic the numerical closure approximation induced by the lifting operator.  Finally,  we analyze the dynamics generated by these algorithms for larger and larger coarse time--steps (i.e., increasing the time between liftings), we will discuss the effects and limitations of the "healing principle'' in this context (Section~\ref{sec:healing}). 

\subsubsection{Coverage is insufficient as macroscopic state variable}\label{sec:lift_r}

\begin{figure}
 \begin{centering}
\includegraphics[width=0.49\textwidth]{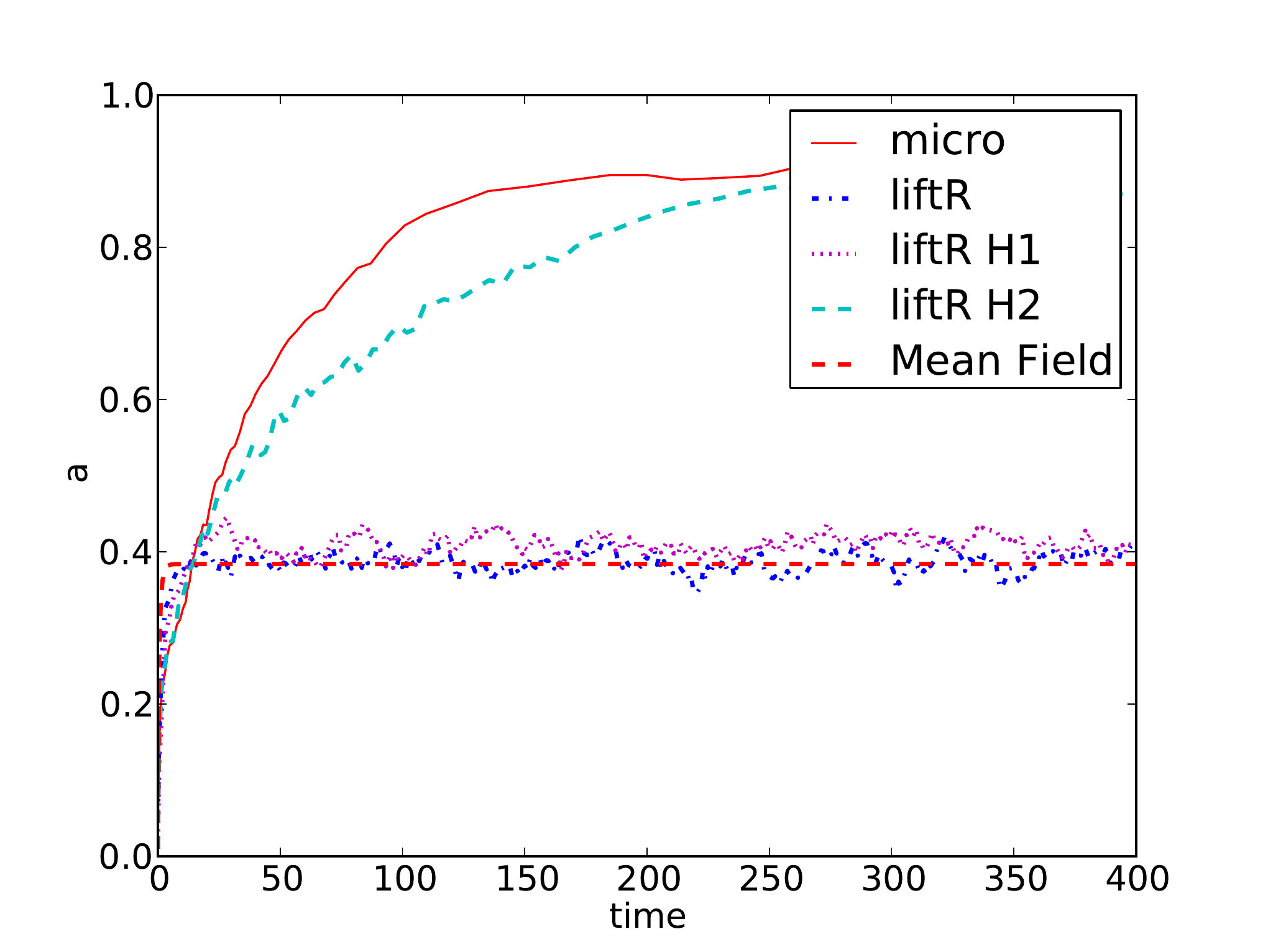}
\includegraphics[width=0.49\textwidth]{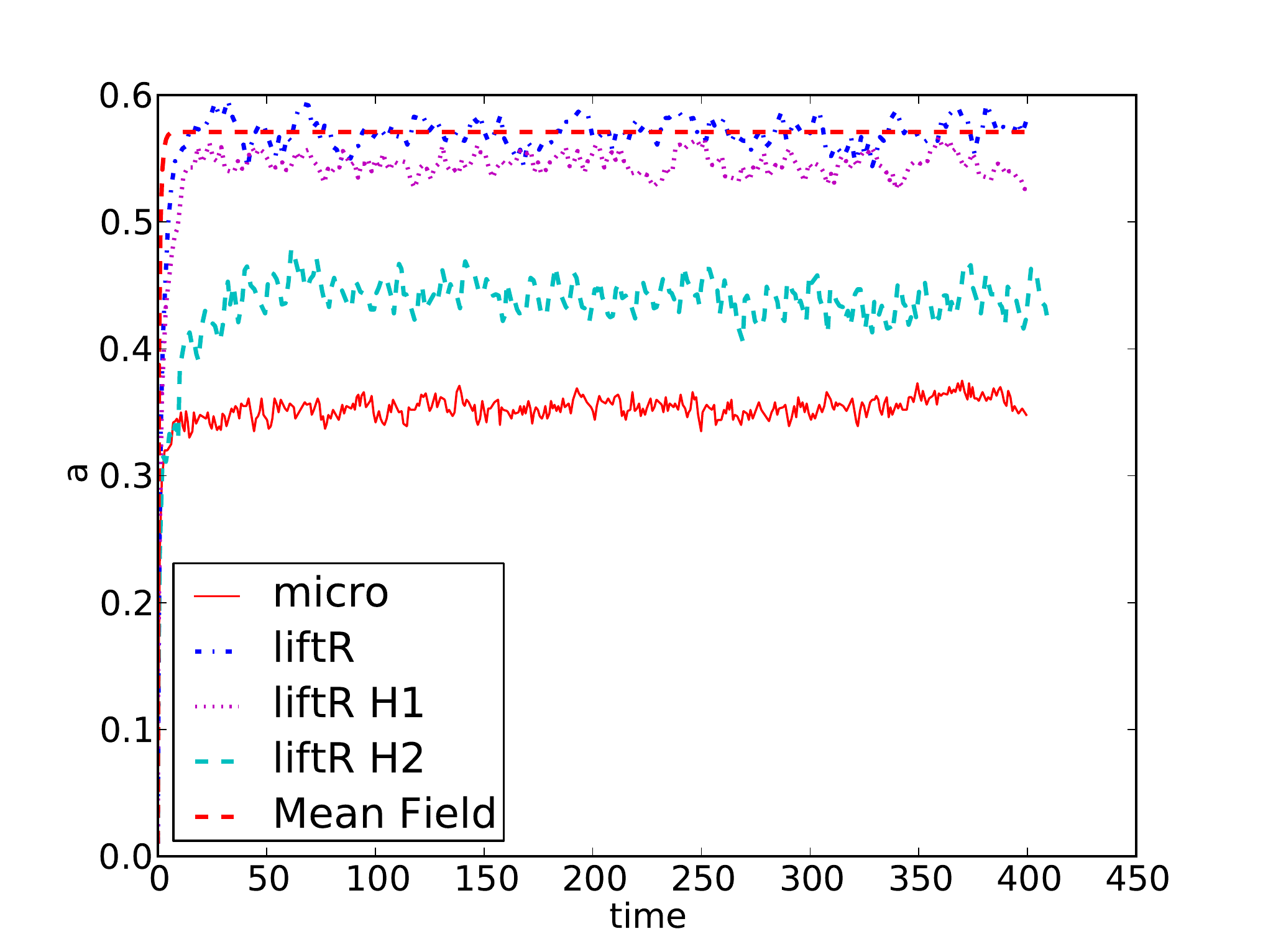}
\caption{\label{fig:healing}Failure of healing, results for {\textbf lift}$_R$ for examples \ref{ex:schl} with $K=K_1$  (up) and with $K=K_2$ (down). The results refer to runs with $1000$ microscopic time steps as coarse time stepper between every lifting/restriction procedure ($H_1$) and with $10000$ microscopic time steps ($H_2$).  Mean Field refers to (\ref{mfsch}). The total simulation time was $~300000$ steps.}
 \end{centering}
 \end{figure}

We first present results using the coverage $a$ as the only macroscopic state variable, from which we lift using the procedure \textbf{lift}$_{R}$, see Remark~\ref{rem:lift_r}.  The numerical results, shown in Figure~\ref{fig:healing}, indicate a significant deviation of the dynamics, as well as of the obtained equilibrium.  In fact, in this case the lifting simply leads to the corresponding 
mean field dynamics. This result is not surprising per se:  when taking into account only the coverages, and randomly placing the elements of species $A$, one is in fact stirring the system, so spatial correlations are lost, and the system effectively satisfies the requirements to display mean field behavior. 

In the literature \cite{makeev} it has been argued that errors in the macroscopic evolution that are induced by an incorrect microscopic initial condition may be able to ''heal"  by taking a larger coarse  time step $\delta t$.
Therefore, we enlarge the size of the coarse time step (to allow for healing).
Figure \ref{fig:healing} shows that this leads to an improved agreement. However, for our system, the healing effect does not suffice for good macroscopic evolution. In our numerical tests we see that performing few hundreds of lifting procedures in a long (hundreds of thousands steps) simulation is sufficient to spoil the microscopic dynamics and to modify the equilibrium values.

Obviously, using only the coverages as macroscopic variables does not seem to be, generally speaking, a good starting point for lattice reactive systems.

\subsubsection{Selection of macroscopic state variables}\label{sec:macro_state}

\begin{figure}
 \begin{centering}
\includegraphics[width=0.49\textwidth]{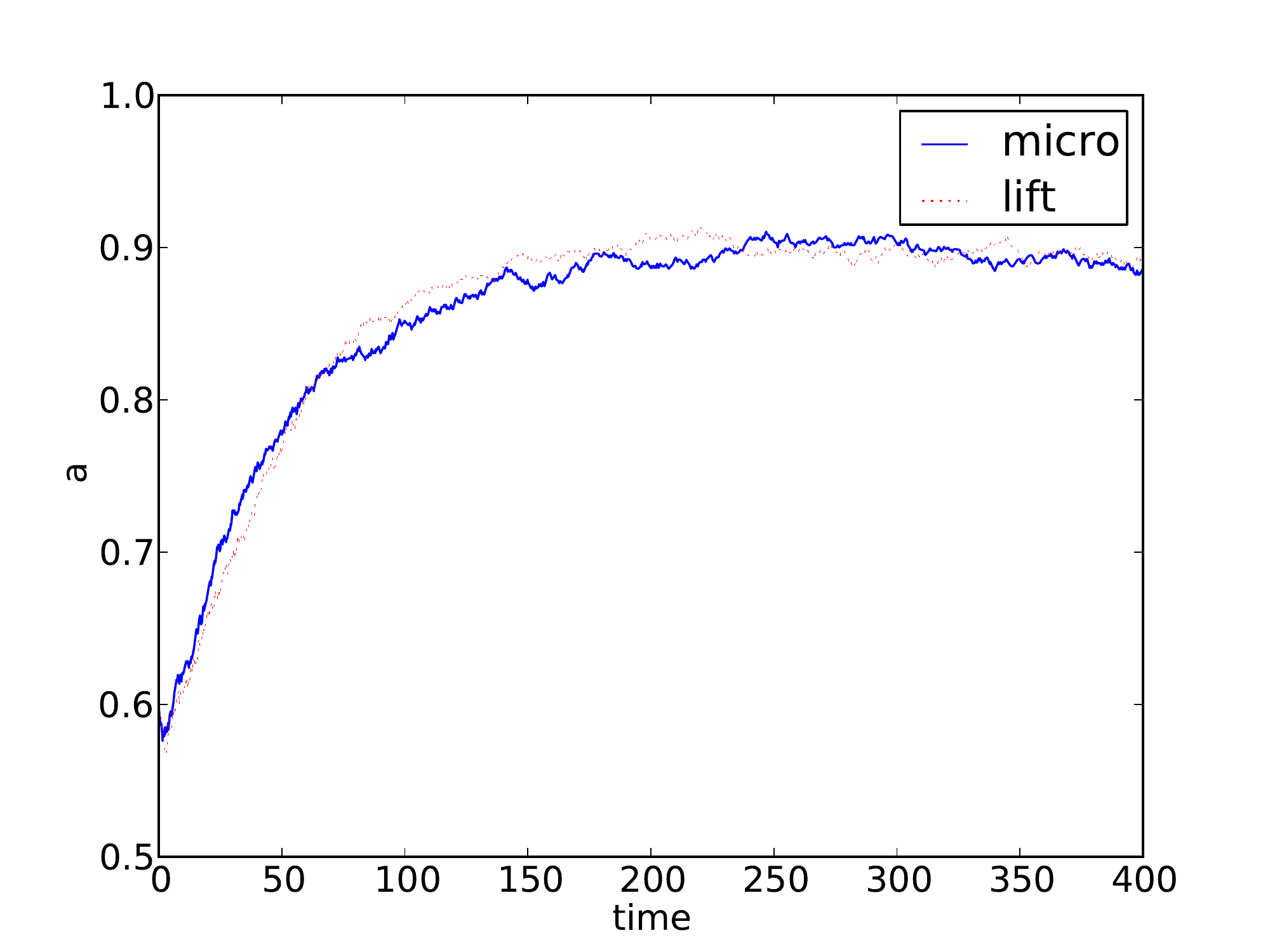}
\includegraphics[width=0.49\textwidth]{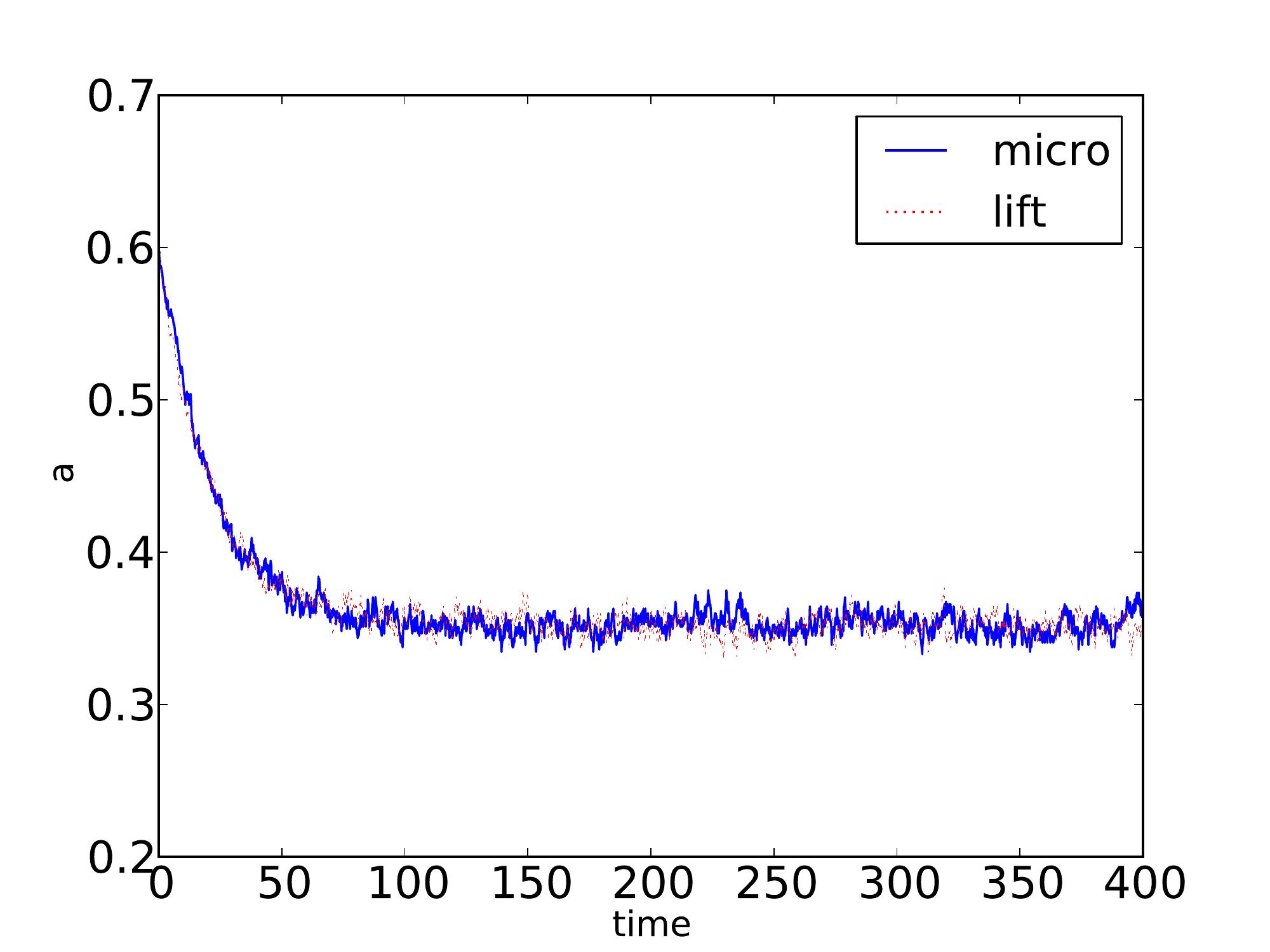}
\caption{ \label{fig:lx100}Comparison of microscopic (continuous line) and restriction/lifting (dashed line) runs with $L_A=100=L_X=100$. The results shown are for reaction rate $K_1$ (up) and $K_2$ (down). The runs refer to {\bf lift}$_B$, but equivalent results are obtained for the other lifting methods.}
 \end{centering}
 \end{figure}

Given the results from the previous section, we augment the set of macroscopic state variables which in turn induces the necessity to choose a more sophisticated lifting procedure (as discussed above). As a first sanity check for the methodology, we initially perform a simulation with the coarse time-stepper using a very high number of macroscopic state variables, i.e., $L_X=L_A=N/20$.
Obviously, in this case we perform almost no reduction, in the sense that only a few microscopic configurations are consistent with the considered macroscopic state.  Figure~\ref{fig:lx100} shows the results for both selected sets of reaction rates in the case of for {\bf lift}$_B$, equivalent results were obtained with the other lifting procedures but are not shown here. As expected, the agreement with the full microscopic simulations is almost perfect in this case.
Our objective here is however to find the simplest (and most effective) algorithm that correctly reproduces these simulations.
In this context, one will want to determine and use the smallest possible set of macroscopic variables for 
the corresponding coarse time-stepper. We propose to base this choice on the previously introduced dynamical equivalence,
which in our sense is a good measure of the instaneous ``reactivity'' of a system.  

Remember that we have considered two microscopic configurations to be equivalent if in both configurations, every reaction in the system is equally likely, i.e., in both configurations, there is an equal amount of reactive elementary blocks, see equation~\eqref{eq:equiv}. For the chemical reactions~(\ref{kin_law_schlogl}), we observe that the first reaction introduces a dependence on the neighboring sites, while the second one does not. Consequently, when choosing $L_A$ and $L_X$ we focus on preserving the number of possible reactants of the first reaction, as the possible reactants for the second reaction are already determined by the coverages. 

Let us first decide upon $L_A$. 
In particular, we note that isolated $A$s  are the only possible elements which may be modified by the first reaction. Therefore, we choose $L_A=1$. The choice of $L_X$ is more difficult, as one can never preserve the number of elements $X$ which are surrounded by $X$s, unless $L_X$ is very large. However, we anticipate (and will show later on) that by keeping $L_A=1$, while increasing $L_X$, we should be able to get to the smallest set of macroscopic variables for which the coarse time-stepper recovers the macroscopic dynamics of the original microscopic system.
For small values of $L_X$, we will show that the quality of the recovered macroscopic dynamics depends on the specifics of the lifting operator.

\subsubsection{Lifting technique lift$_A$}\label{sec:lift_A}

\begin{figure}
 \begin{centering}
\includegraphics[width=0.49\textwidth]{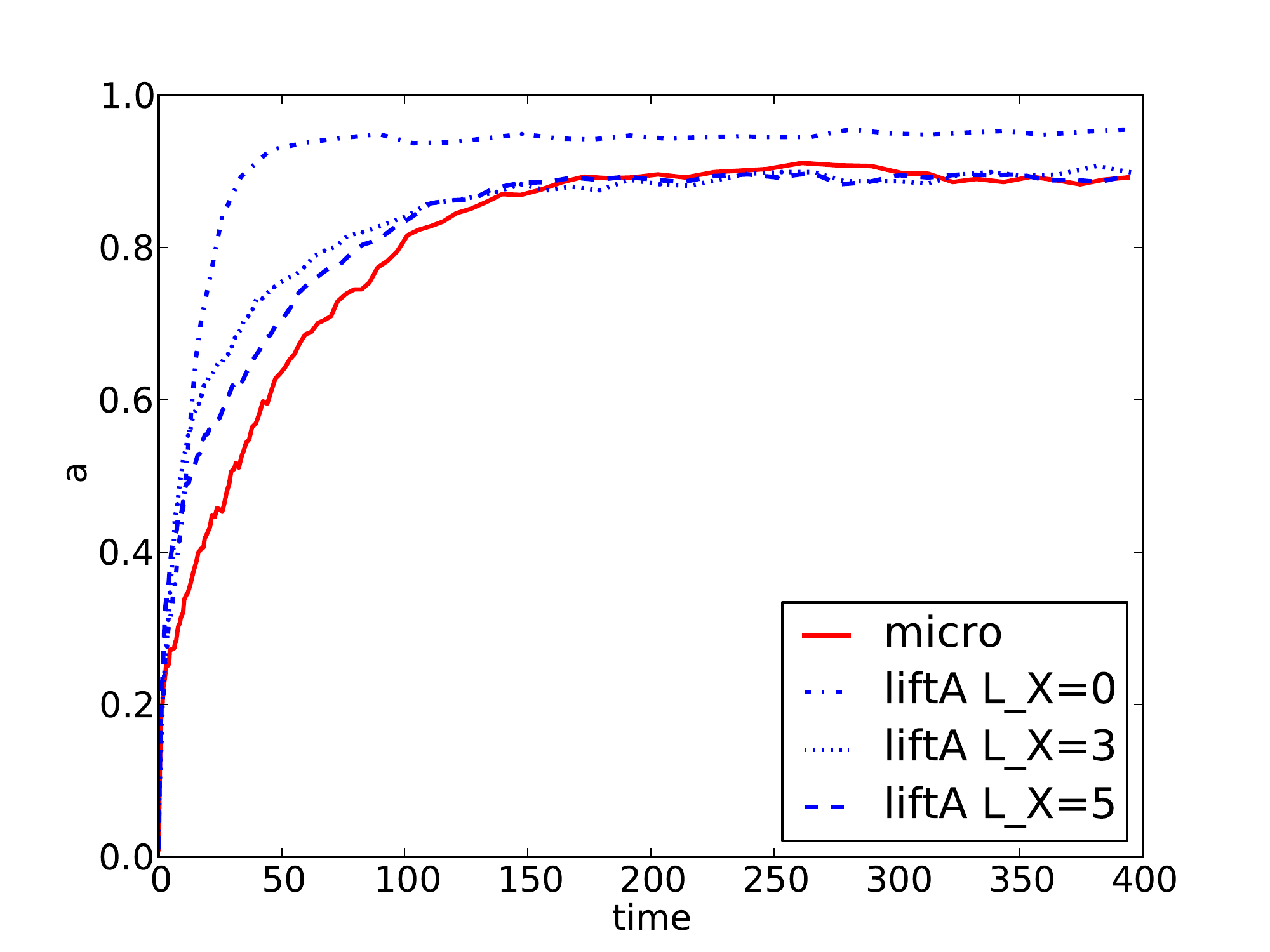}
\includegraphics[width=0.49\textwidth]{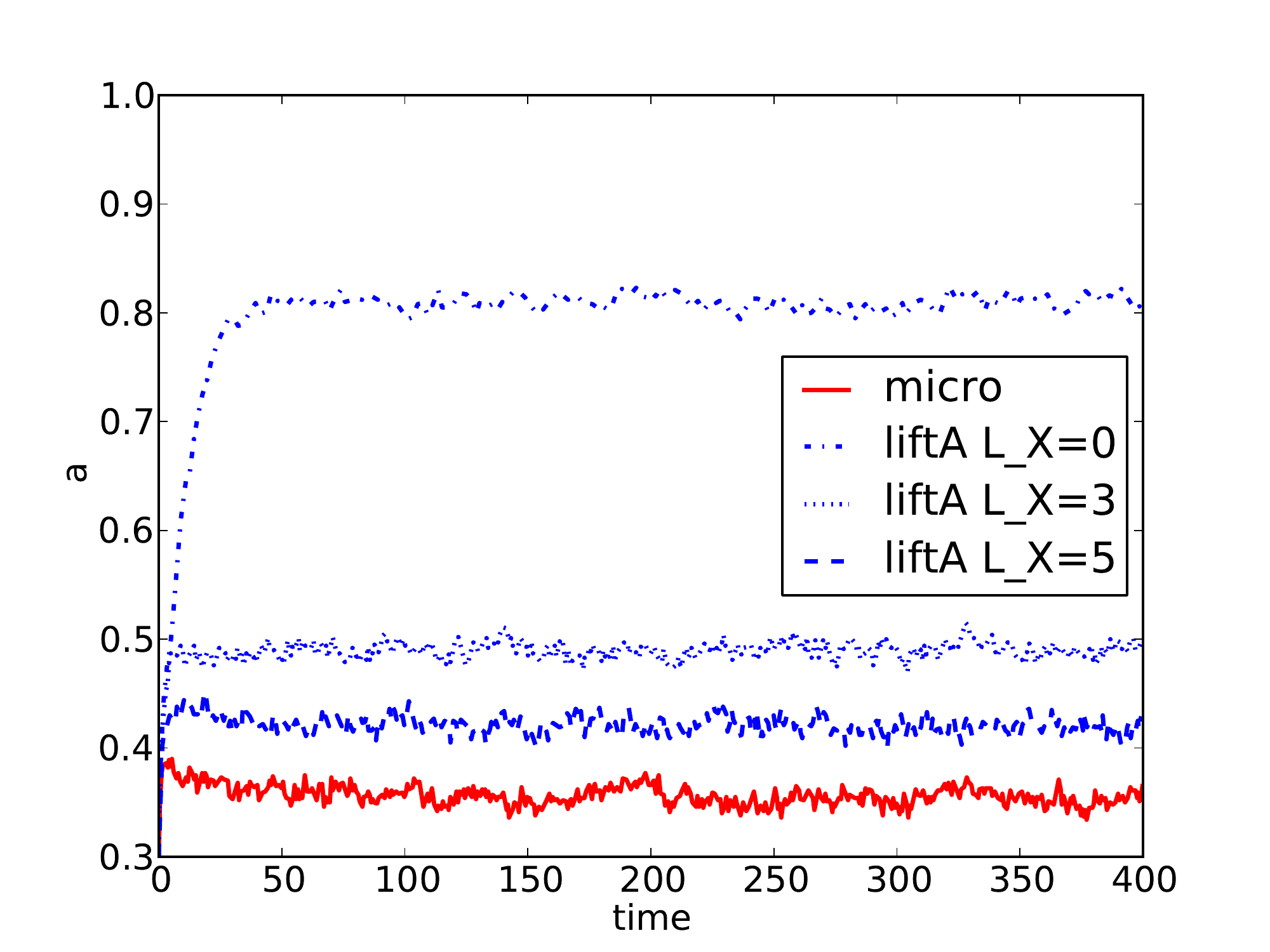}
\caption{ \label{fig:liftA}Restriction/lifting results for {\textbf lift}$_A$ for both sets of reaction rates $K=K_1$ (up) and $K=K_2$ (down). Comparison between different choices of $L_X$.}
 \end{centering}
 \end{figure}

We now proceed by performing a simulation using the coarse time-stepper with the lifting \textbf{lift}$_A$, using $L_A=1$ and different values of  $L_X$. The results for both sets of reaction rates ($K_1$ and $K_2$) are shown in Figure~\ref{fig:liftA} for the smallest possible coarse time step.
For the set of reaction rates $K_1$, when increasing $L_X$, we quickly obtain an excellent agreement between the numerical closure approximation and the fully microscopic simulation. For the set of reaction rates $K_2$, however, we see that for similar values of $L_X$, the macroscopic evolution behaves completely differently (both dynamically and in equilibrium), and we need to increase $L_X$ quite substantially before getting an agreement.

\begin{figure}
\begin{centering}
\includegraphics[width=0.49\textwidth]{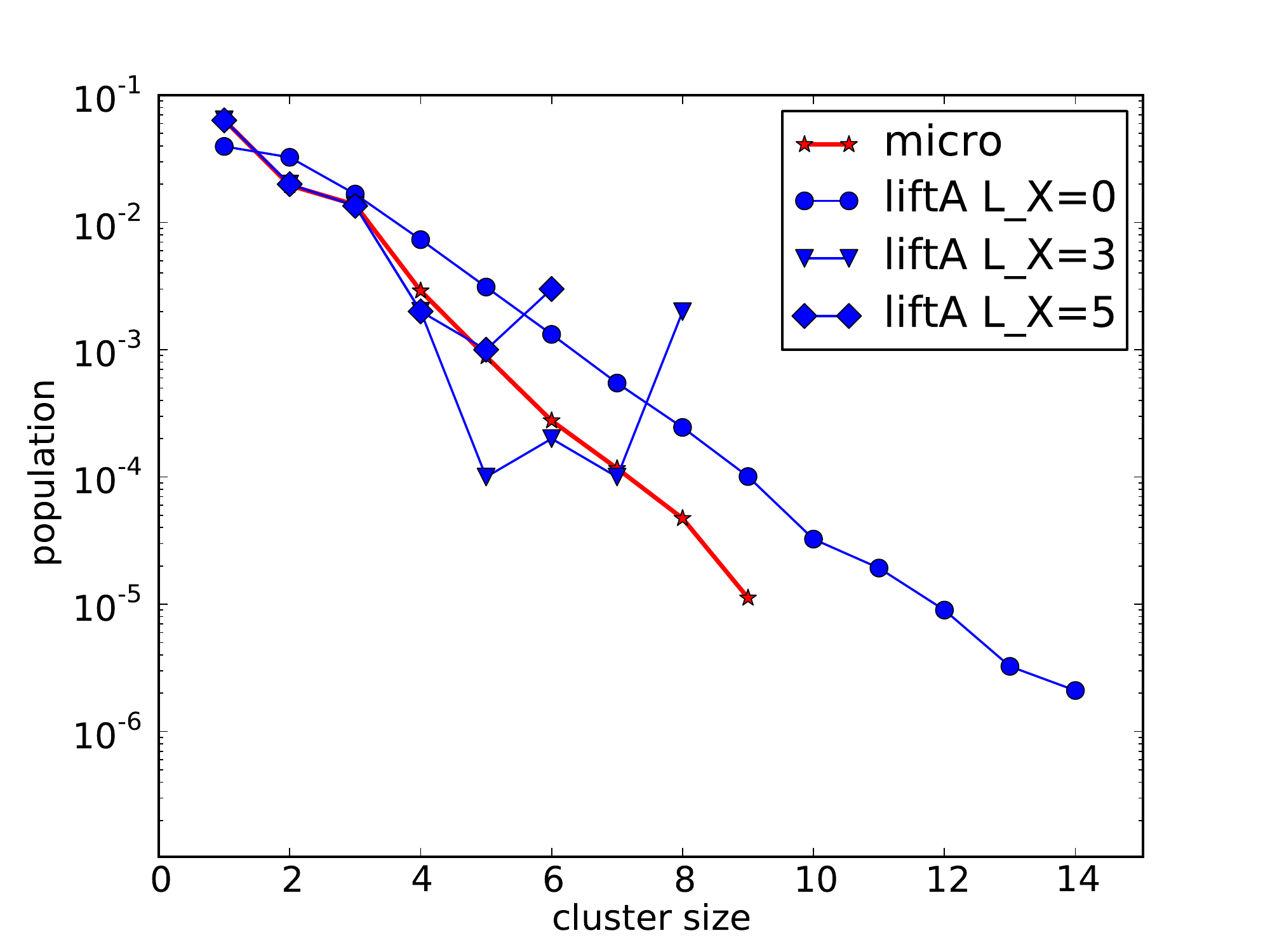}
\includegraphics[width=0.49\textwidth]{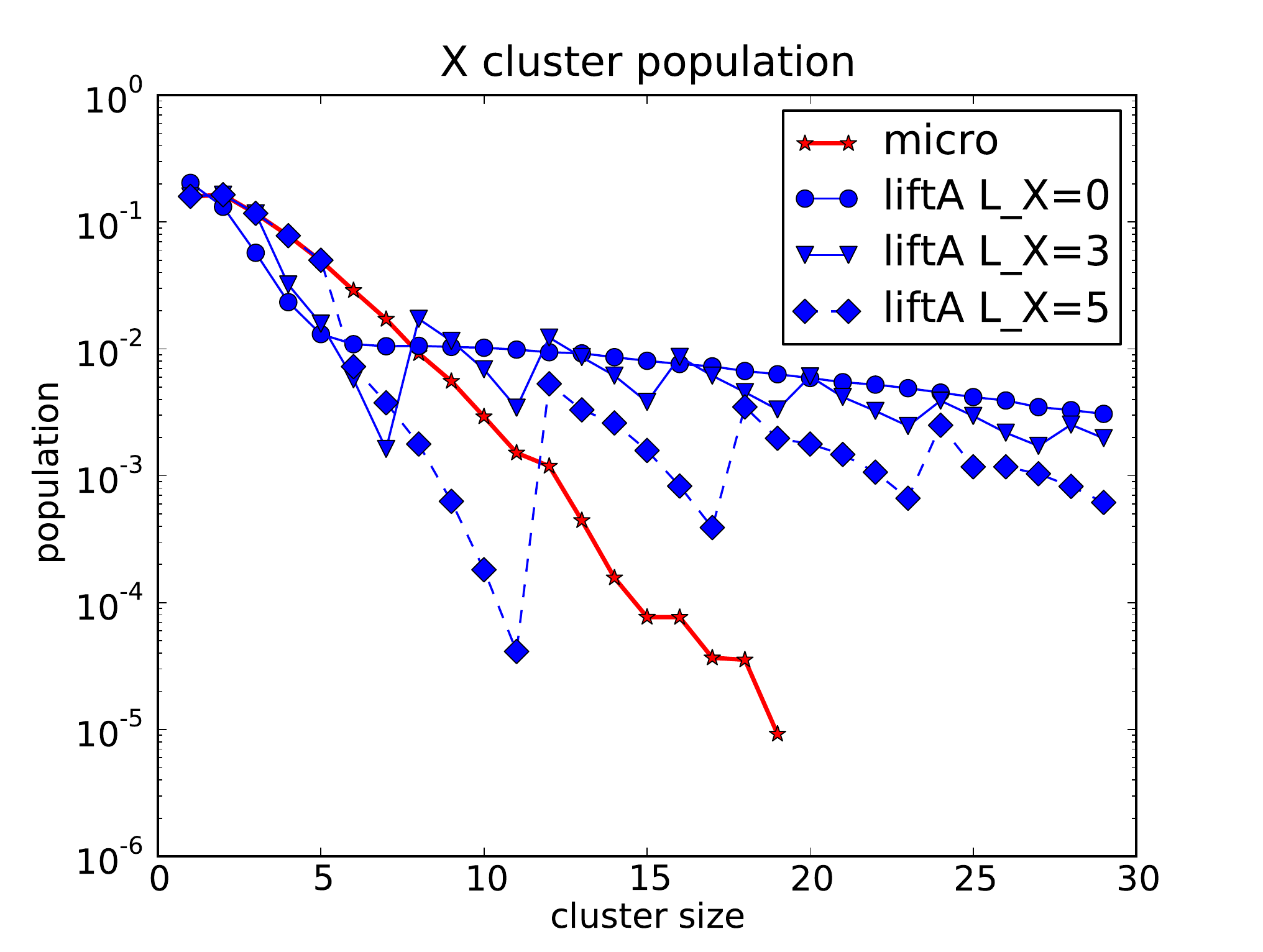}
\caption{\label{fig:loglift}Comparison of cluster populations for species $X$ for {\bf lift}$_A$ at equilibrium in a semilogarithmic scale. The red stars correspond to the average over a microscopic run at equilibrium for $K=K_1$ (up), and $K=K_2$ (down). The other data correspond to the average of $10000$ liftings, given the values of the macroscopic set at equilibrium. }
 \end{centering}
 \end{figure}

This difference can be rationalized by analyzing the distribution of cluster size before and after the lifting.
To this end, we start from an equilibrium microscopic configuration for a ring of length $N=2000$ and count the number of clusters $M^l_A$ and $M^l_X$ ($l=1,\ldots,N$) via equation~(\ref{mxjmaj}) at each time step during a microscopic simulation, for a total time of $T=100$  and average over time,
$$
\overline{M^l_A}=\langle M^l_A(t)\rangle_{\rm eq},\quad \overline{M^l_X}=\langle M^l_X(t)\rangle_{\rm eq},
$$  
where the subscript {\rm eq} emphasizes the fact that the time average is taken over a simulation that is in equilibrium for all values of $t$.
The resulting (normalized) histogram is plotted in Figure~\ref{fig:loglift} (red starred line).  Next, to obtain a probability distribution for the coarse time stepper, we repeat the lifting procedure $10000$ times using method \textbf{lift}$_A$, starting from the macroscopic state variables $M^1_A$ and $M_X^l$ ($1\le l \le L_X$), for different values of $L_X$, and average again the number of clusters of length $1\le l \le N$ over the ensemble of microscopic configurations (see Figure~\ref{fig:loglift}). 

For the original microdynamics, this distribution decays exponentially with the length of the clusters. Such rapid decrease can be explained by observing the reaction laws (\ref{kin_law_schlogl}). Any long sequence of $X$ is likely to be broken in two smaller ones by both reactions, while sequences of $A$s can be interrupted only by the second reaction, so the dynamics greatly favors short clusters of $X$s rather than longer ones. This feature is not preserved by the lifting procedure; as can be seen in Figure~\ref{fig:loglift}, for the lifted distributions, longer clusters are much more likely than for the micro-dynamics. Moreover, we also observe oscillations in the different cluster populations depending on the choice of $L_X$ which determine the size of the blocks that are contained in the remainder $B^r_X$.

While appealing due to its simplicity, the lift$_A$ procedure thus suffers from some major drawbacks that can be related to its inability to correctly preserve the distribution of clusters size. 
It is relevant at this stage to identify the reasons behind this inaccuracy.
Recall that in \textbf{lift}$_A$,  the idea is to maximize the number of clusters of species $A$ and $X$ created with the remainders, but that no special attention is being paid to the relative values of $C_A$ and $C_X$. As a result, in the final step of the
lifting, multiple building blocks of the same species can be selected,
resulting in longer clusters. To analyze the lifted distributions, we need to
estimate the probability of picking in sequence $k$ blocks corresponding to
species $X$ from $B_X^r\cup B_A^r$ (see end of Section~\ref{sec:lift_proc}).
For small values of $k$, relative to the total number of available blocks,
i.e., $k\ll n^r$, this probability can be approximated by a simple multiplication of  the probability of picking first a block of $A$ then $k$ blocks $X$ and then another block $X$: 
\begin{equation}\label{prob_clust}
\mathcal{P}_k\simeq p_X^k (1-p_X)^2 n^r,
\end{equation} 
where $p_X=\frac{C_X}{n^r}$ is the probability of getting a block of $X$s out of the whole set $B_X^r\cup B_X^r$.
Using (\ref{prob_clust}), and considering that the cluster population is simply $k \mathcal{P}_k$, we see that if $p_{X}\simeq 0.5$, i.e. we have approximately equally many building blocks for both species, then we get almost a plateau in the populations for clusters of length slightly larger than $L_X$. In Figure~\ref{fig:loglift} we see such a plateau when $L_X=0$. Remember that in this case, all elements of species $X$ are contained in the remainder $B_X^r$, as the set $B_X$ is empty.  When choosing $L_X>0$, the behavior is different. In that case, we observe oscillations, which can be explained by noting that clusters are built by concatenation of building blocks, and hence, multiples of $L_X+1$ are more likely to occur.

Increasing $L_X$ gives better results, because the number of building blocks of species $X$ in the remainder diminishes, and, moreover, the size of the those blocks also increases.

\subsubsection{Lifting technique lift$_B$}\label{sec:lift_B}

\begin{figure}

 \begin{centering}
 \vspace{-0.5cm}
\includegraphics[width=0.49\textwidth]{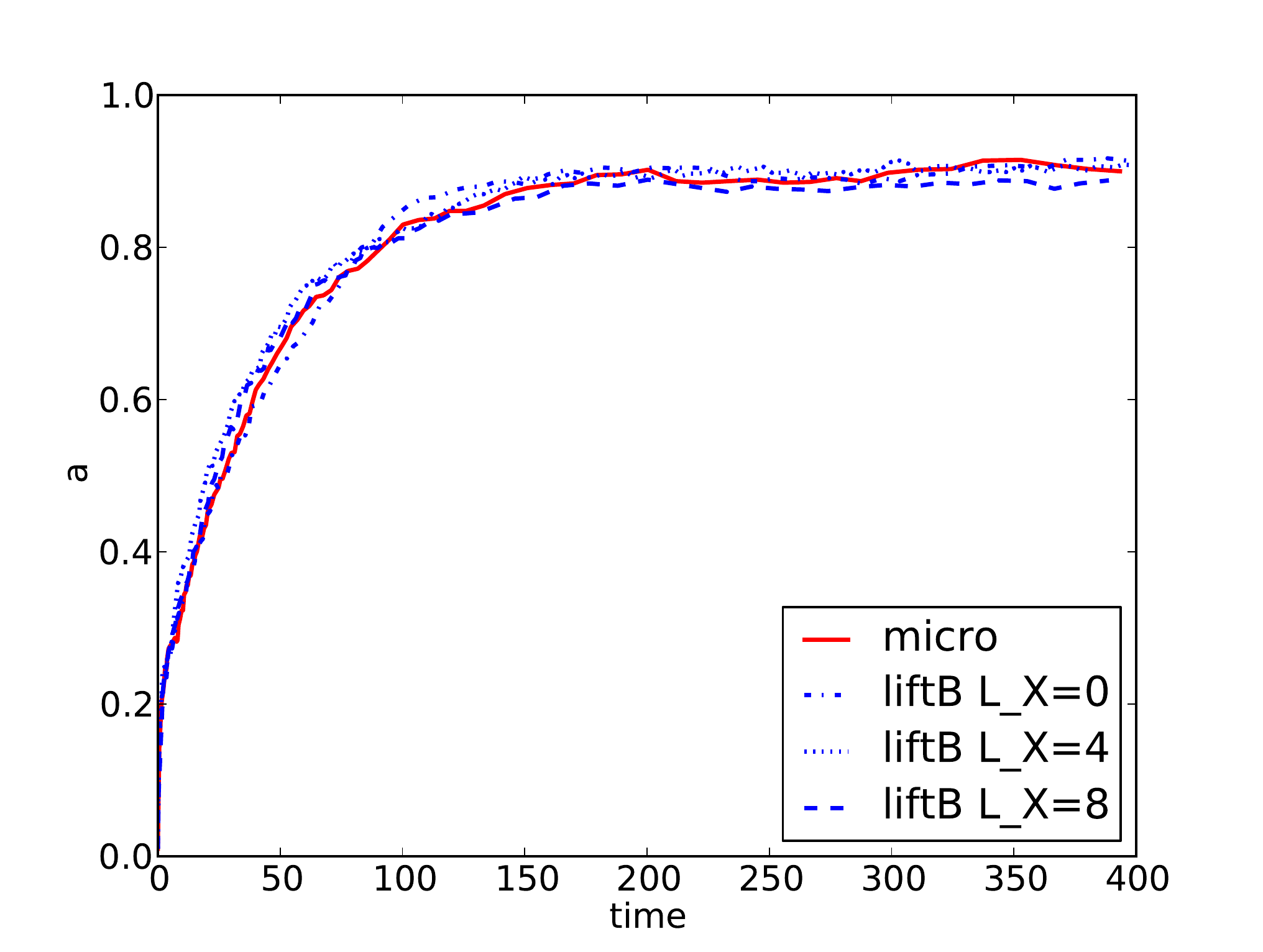}
\includegraphics[width=0.49\textwidth]{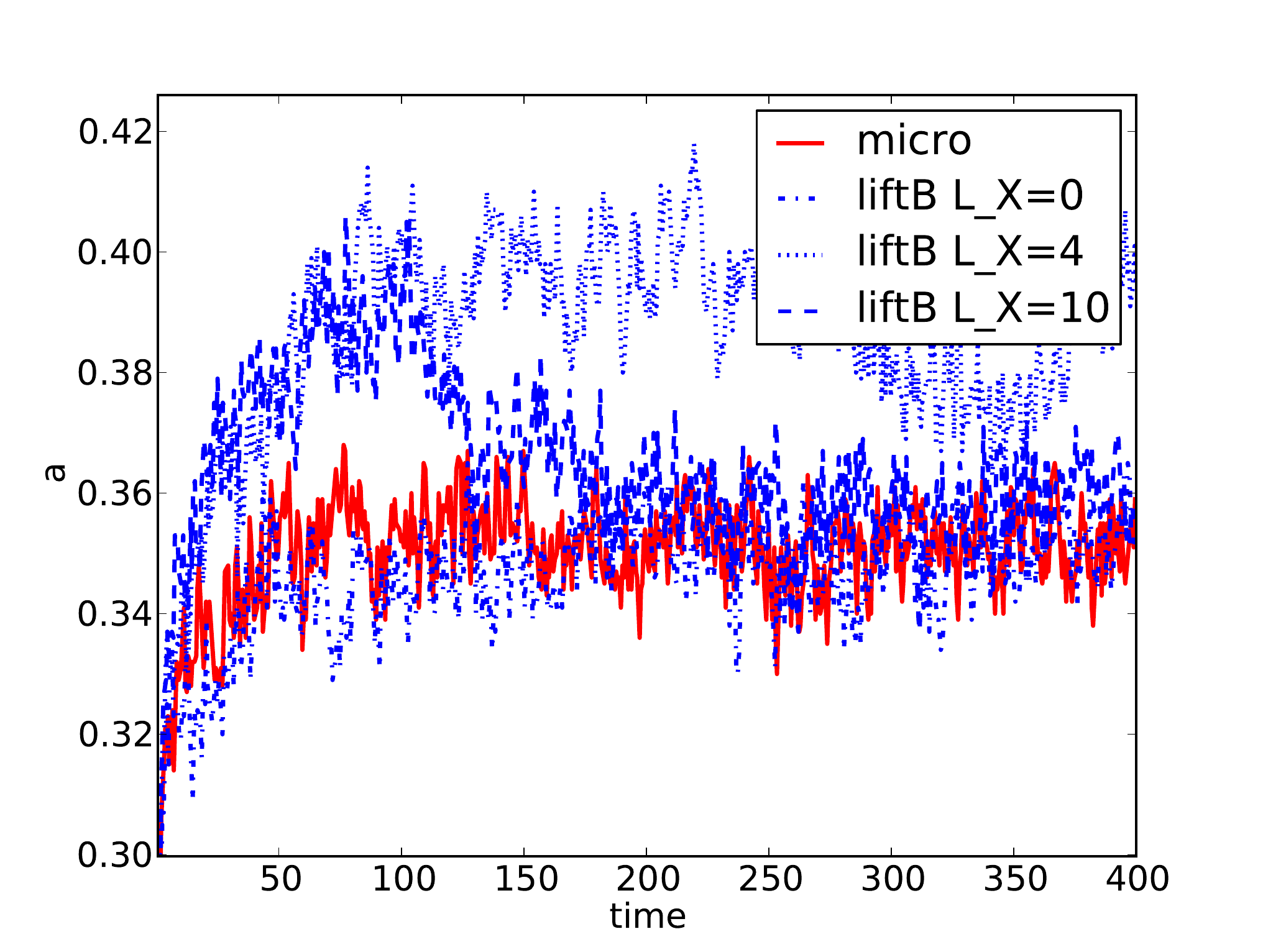}
\caption{ \label{fig:lift2}Restriction/lifting results for {\bf lift}$_B$ for both sets of reaction rates $K_1$ (up) and $K_2$ (down); the figure on the right is zoomed to the equilibrium region.}
 \end{centering}
 \end{figure}

We now repeat the previous experiment for the lifting operator \textbf{lift}$_B$, thus starting with the minimum coarse time step. For the reaction rates $K_1$, we observe a much better agreement with the microscopic data for small values of $L_X$. Choosing $L_X=1$ already gives excellent agreement, see Figure~\ref{fig:lift2} (left).
For $K_2$, instead, the results are at first glance somewhat surprising. In fact, we get a very good agreement between the coarse time-stepper and  the microscopic simulation by using only $L_X=0$ and $L_A=1$. However, when we add extra macroscopic variables, the results initially worsen.

The main reason for this unanticipated behavior may be found by analyzing the
dependence of the minimal and maximal possible number of clusters in the
remainders ($C^{\min}_{\{X,A\}}$ and $C^{\max}_{\{X,A\}}$, see equations
\eqref{maxC} and \eqref{minC}) on $L_X$ and $L_A$.
We observe that a number $m_A-m_X$ of blocks of species $A$ may be left over after placing all elements contained in $B_X$. These elements need to be placed in the final step of the lifting, together with the remainders of species $X$. For fixed $L_A$, the number $m_A-m_X$ reaches its maximum at $L_X=0$, so $C^{\min}_X$ is maximal for $L_X=0$. Simultaneously, because of the fact that $B^r_X$ then contains all the elements of species $X$  present in the system, $C^{\max}_X$ is maximal for $L_X=0$. A high value for $C^{\min,\max}_X$ implies that the elements of species $X$ will be distributed in a large number of short clusters, which corresponds (by coincidence) to the structure of the microscopic realizations. 

\begin{figure}
\begin{centering}
\vspace{-0.5cm}
\includegraphics[width=0.49\textwidth]{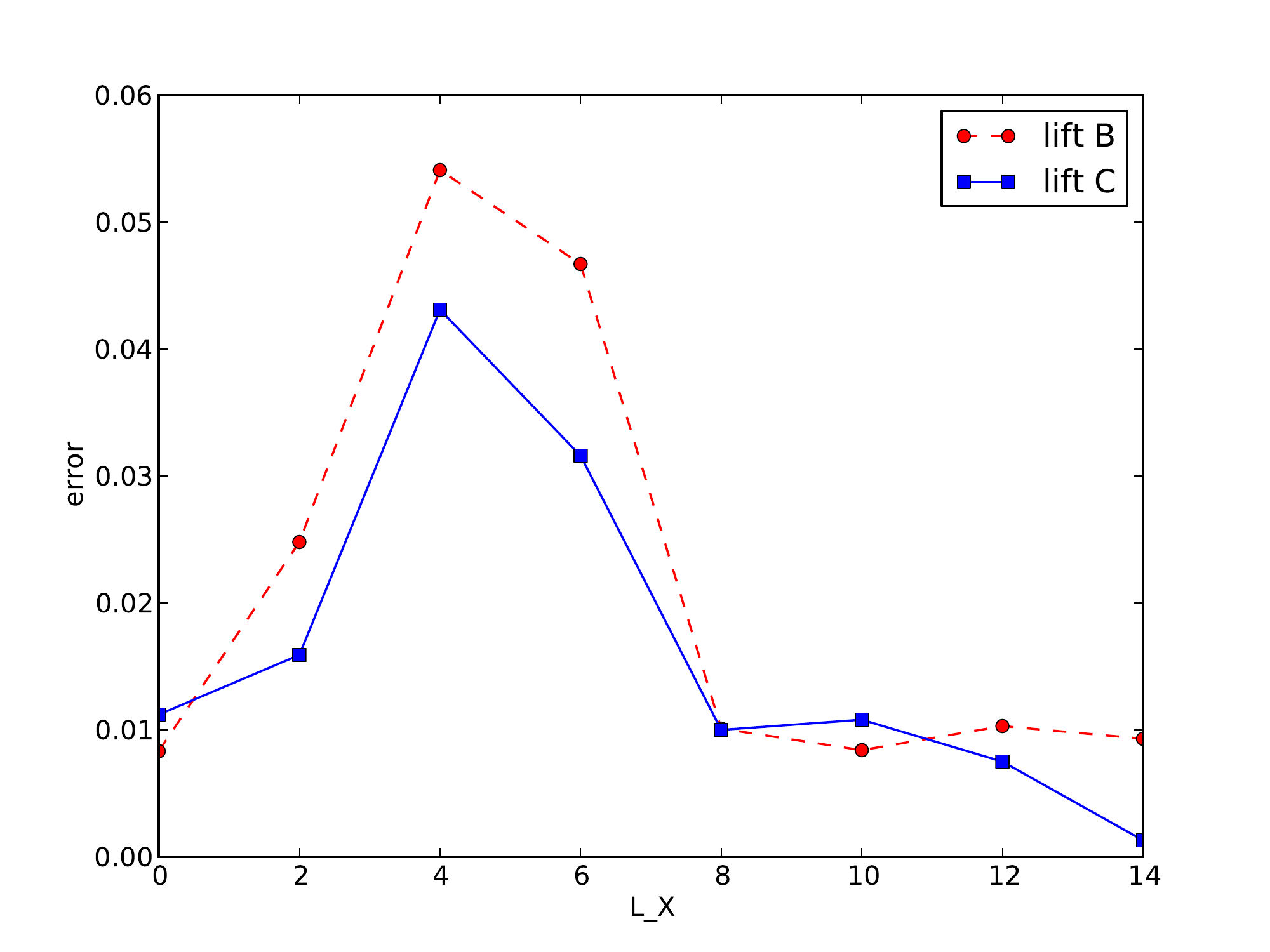} 
  \caption{\label{fig:err_lx} Effect of increasing $L_X$ using {\bf lift}$_B$ and {\bf lift}$_C$ when $K=K_2$. The error represent the time average of the square difference between the $a$ coverage data coming from a microscopic run and a lifting/restriction one with initial conditions at equilibrium.} 
    \end{centering}
\end{figure}

\begin{figure}
 \begin{centering}
\includegraphics[width=0.49\textwidth]{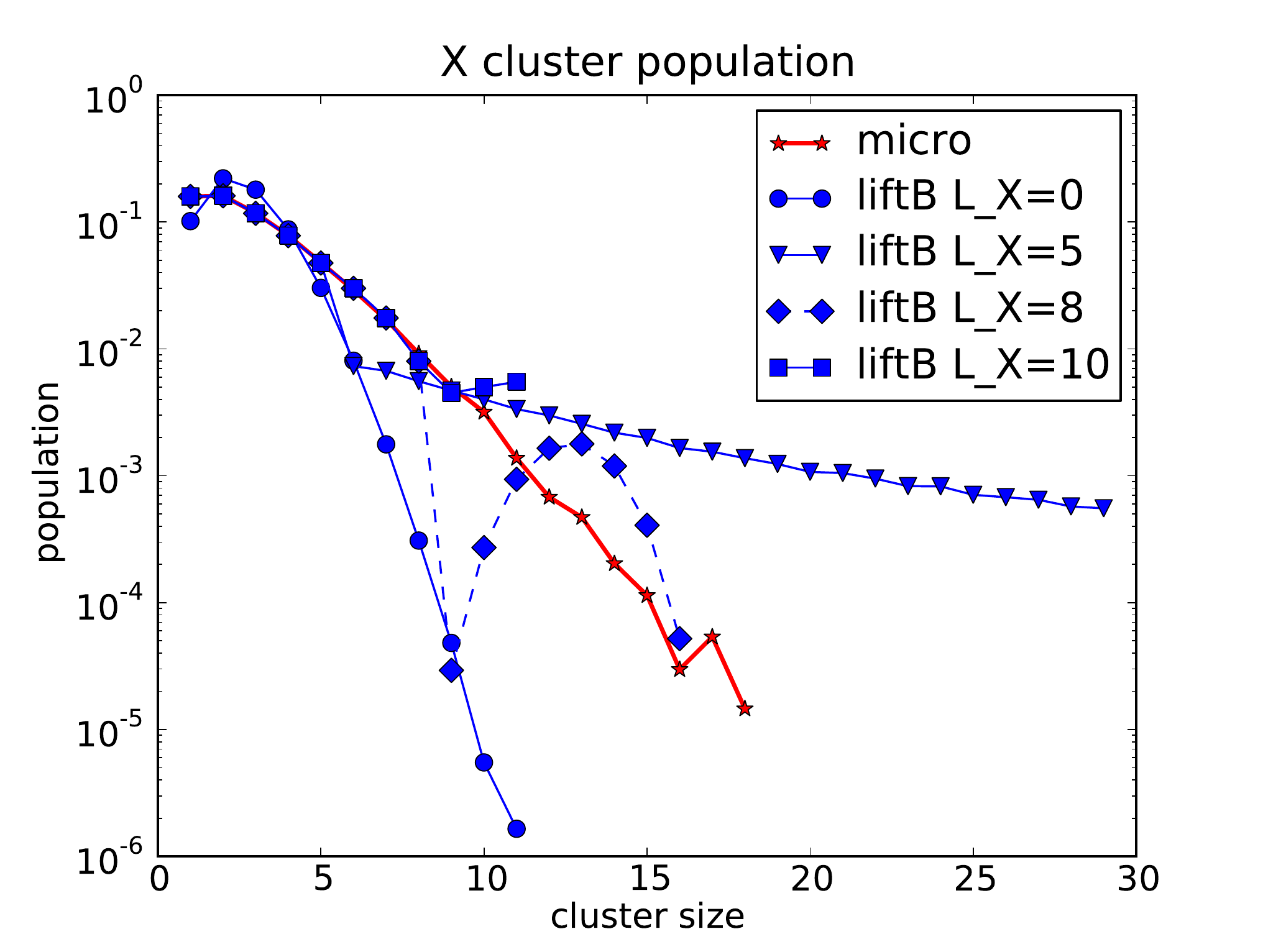}
\includegraphics[width=0.49\textwidth]{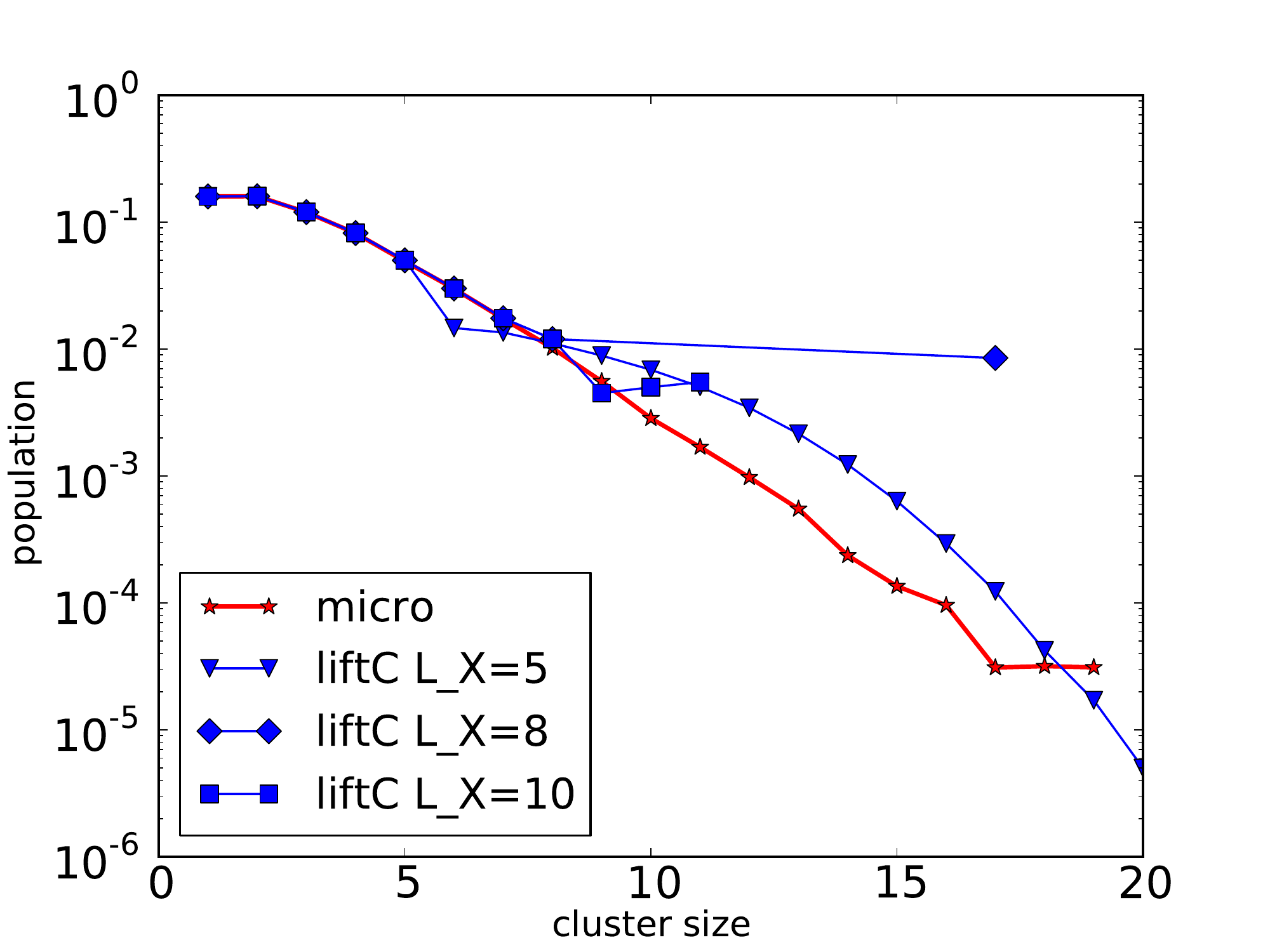}
\caption{\label{fig:clustBC}Comparison of $X$ cluster population for \textbf{lift}$_B$ (up) and \textbf{lift}$_C$ (down) at equilibrium in a semilogarithmic scale, the micro data refer to the average over a microscopic run at equilibrium for $K_2$, while the lifting are the repetition of $10000$ liftings given the values of the macroscopic set at equilibrium. }
 \end{centering}
 \end{figure}

When  $L_X>0$, the quantity $m_A-m_X$ decreases, and so does $C^{\min}_X$,  until it reaches its minimum value of $1$, corresponding to a case in which all remaining elements of species $X$ are contained in a single long cluster. (Note that, theoretically, one may also encounter situations in which the set $B_X^r$ is empty, corresponding to $C_X=0$.) It is then that we observe the largest error for the lifting. After this peak, the lifting performs better, as the number of elements in $B_X^r$ decreases, leaving less room for artifacts in the evolution (Figure~\ref{fig:lift2} and  Figure\ref{fig:err_lx} ).
As for \textbf{lift}$_A$, we have evaluated the $X$ cluster population at the equilibrium for $K=K_2$ and the results, shown in Figure~\ref{fig:clustBC} (left), are in line with the above observations. We note that the decay in cluster populations as a function of cluster length now becomes non-oscillatory. Nevertheless, this decay is still slow compared to that observed in the full microscopic simulation.

To summarize, when increasing the number of macroscopic variables $L_X$, we have a competition between two opposite effects. On the one hand, as expected, taking into account more variables reduces the degrees of freedom that are affected by the lifting, increasing the accuracy. On the other hand, we observe that, for intermediate values of $L_X$, a bias is introduced (via $C^{\min}_X$ in this case) in the distribution of the reconstructed elements.  Depending on the relative importance of these effects, convergence to the microscopic dynamics may be non-monotonic. Note that this does not explain the very good agreement for the case $L_X=0$, which is coincidental, and appears to be caused by the fact that lifting errors that artificially generate clusters that are too short do not affect the macroscopic dynamics.

\subsubsection{Lifting technique lift$_C$}\label{sec:lift_c}

\begin{figure}

 \begin{centering}
\includegraphics[width=0.49\textwidth]{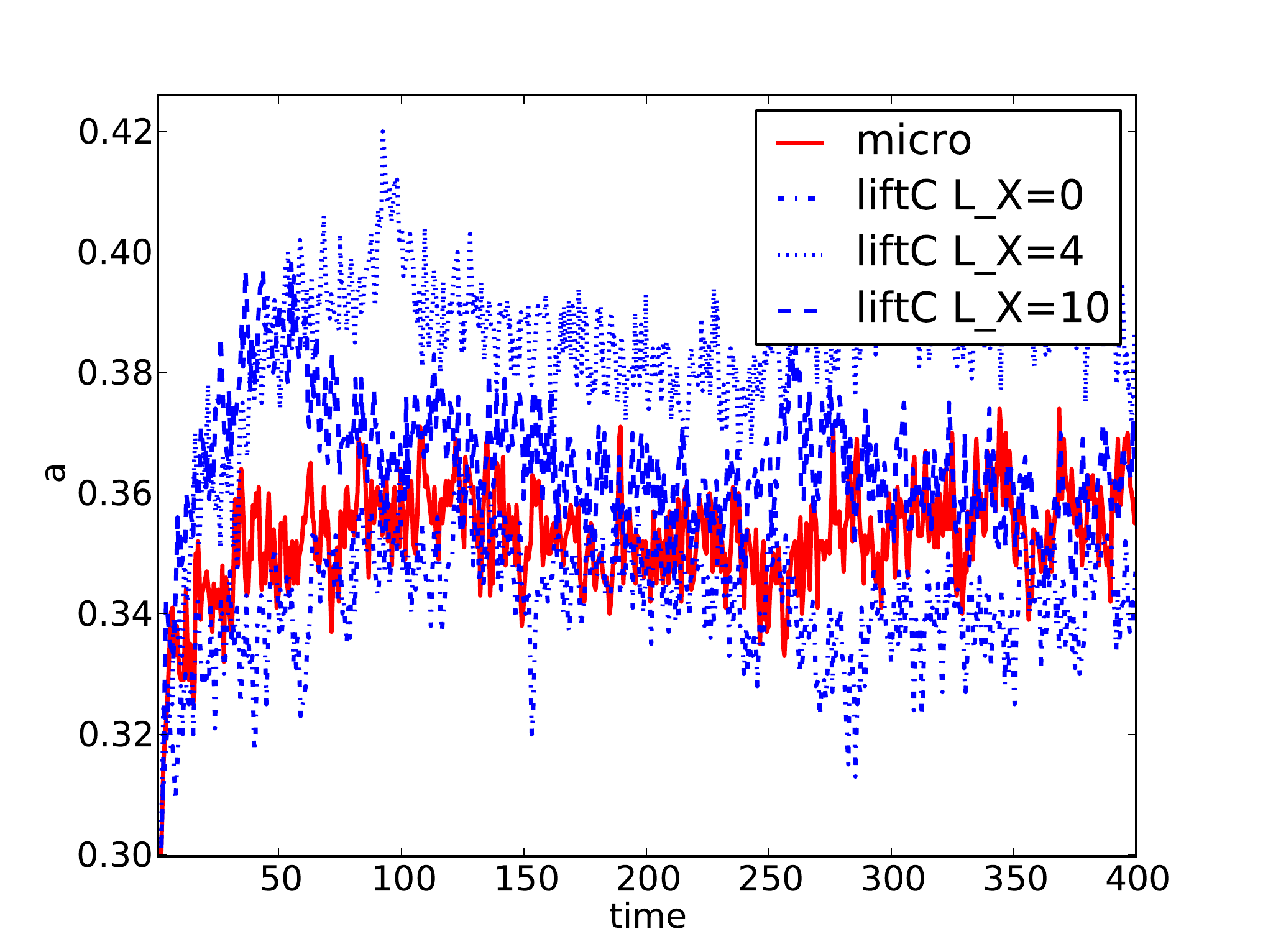}
\caption{ \label{fig:lift3}Restriction/lifting results for {\bf lift}$_C$ and $K=K_2$.}
 \end{centering}
 \end{figure}
 
Finally, we repeat the experiment for the lifting operator \textbf{lift}$_C$.
The idea is here to use an estimate of the cluster size distribution to find an appropriate estimate 
for the minimum number of clusters: we may for example use 
(\ref{prob_clust}) to get an estimate for the maximum length, from which such number can be deduced. 
We set $C^{\min}_X=\tilde{k},$ with $\tilde{k}$ such that
\begin{equation}
p_X^{\tilde{k}} (1-p_X)^2 n^r<\epsilon,
\end{equation} 
for a chosen small value of $\epsilon$, explicitly excluding configurations that are very unlikely from a probabilistic point of view. (Here, we chose $\epsilon=0.03$.) 
In  our numerical experiments, the macroscopic simulation results for the reaction rates $K_1$ were very similar to those obtained with \textbf{lift}$_B$.  This is not surprising, as the lifting technique is similar in the sense that it also picks the number of clusters homegeneously between two imposed values, the only difference between these two schemes being in the estimate of the minimal number of clusters. 
We nevertheless notice that, for $K_2$, the results improve slightly. Considering again the distribution of cluster sizes (see Figure~\ref{fig:clustBC} (right)), we observe that some oscillations on the cluster populations are still present, but overall the distributions obtained with the liftings are closer to the data coming from the microscopic dynamics thanks to a better estimate for one of the boundaries in the number of clusters. 


\subsubsection{Effects of healing}\label{sec:healing}

\begin{figure}
 \begin{centering}
\includegraphics[width=0.49\textwidth]{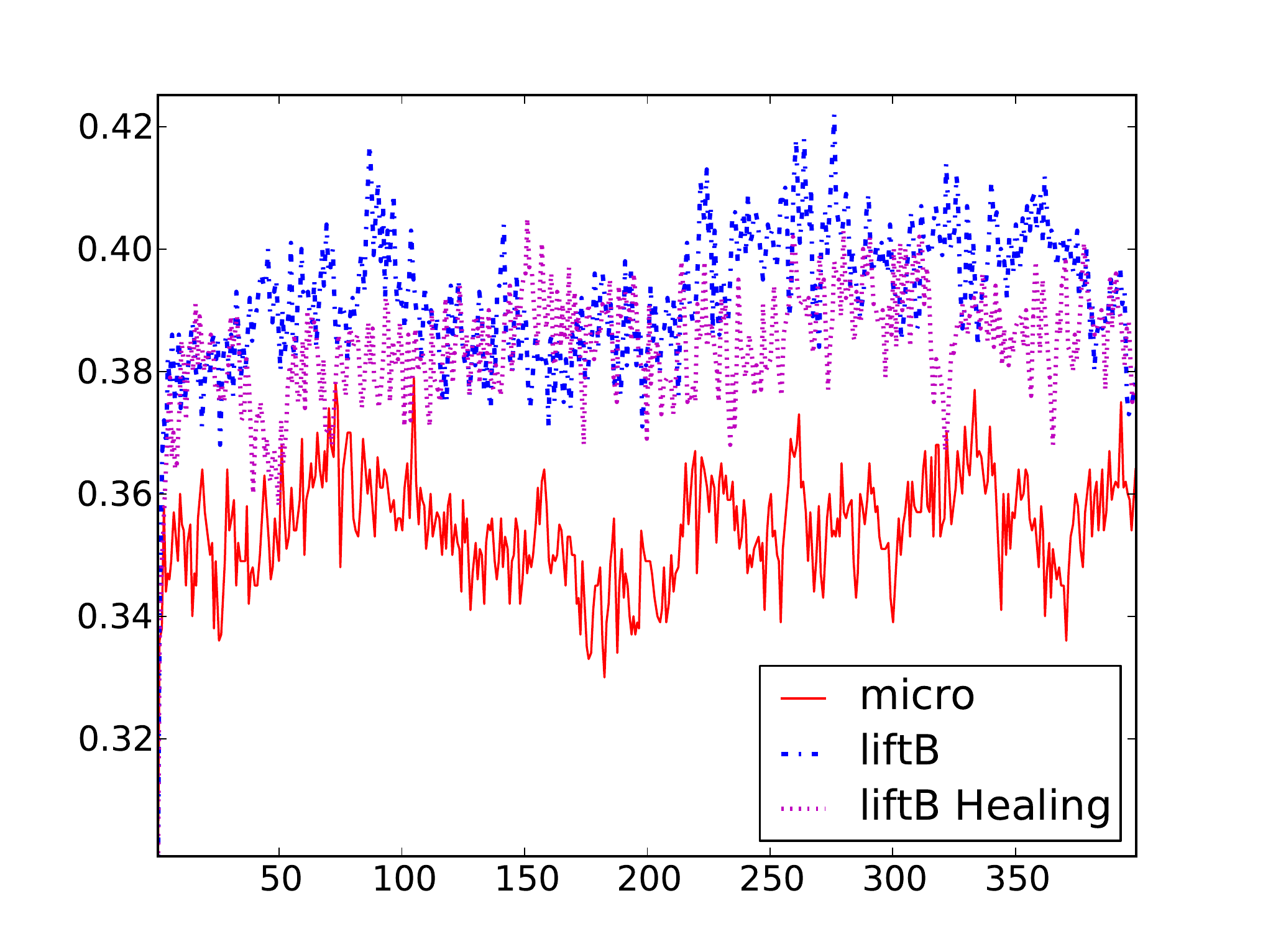}
\includegraphics[width=0.49\textwidth]{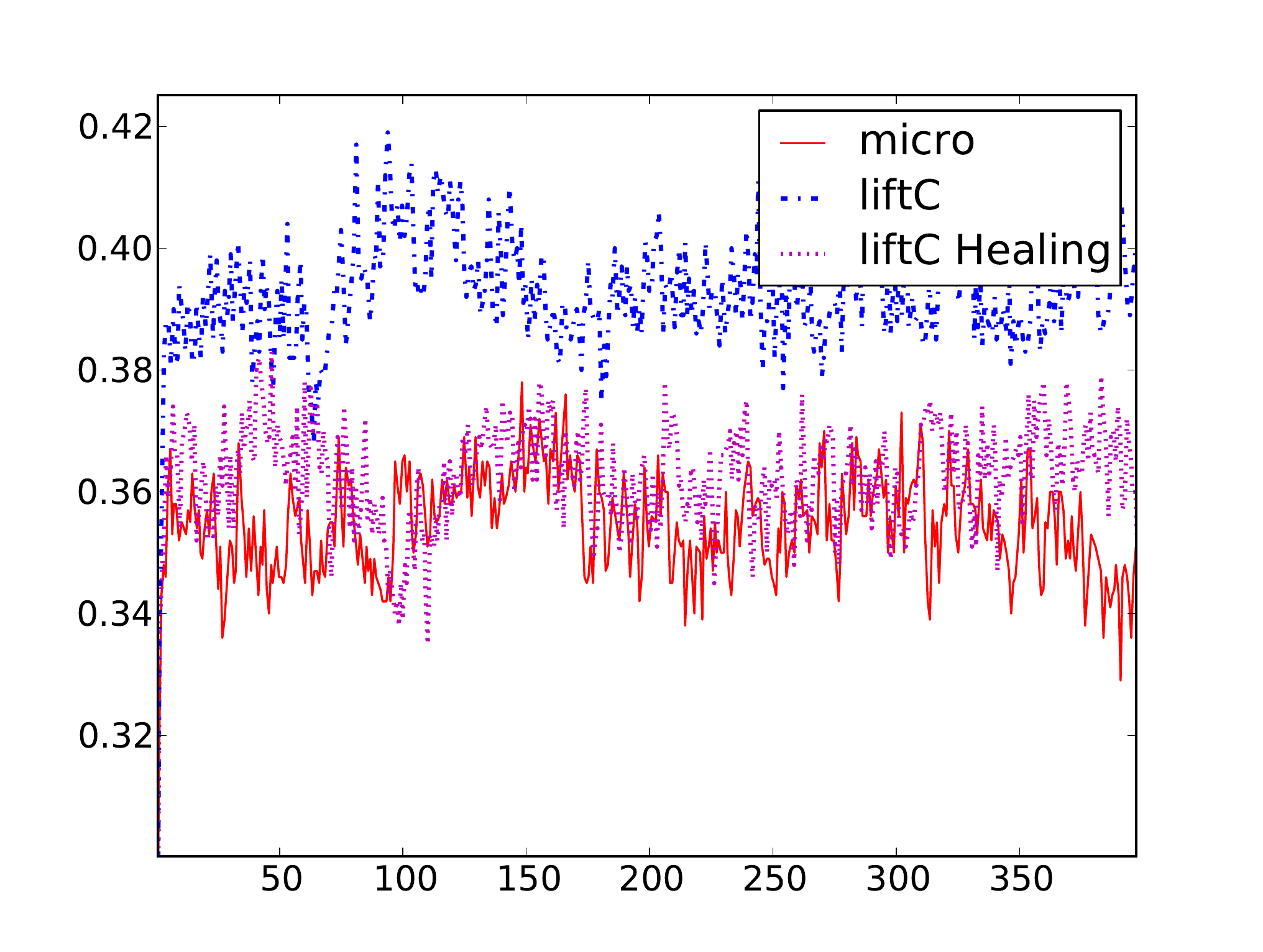}
\caption{ \label{fig:healing_BC}Restriction/lifting results for {\bf lift}$_B$ (up) and {\bf lift}$_C$ (down) for $K=K_2$ with healing. In both the cases we compare the lifting procedure with and without healing (of $1000$ microscopic time steps) The total simulation time is $T=400$, $1000$ steps correspond roughly to a time of $1$. For the parameters we have  $N=3000$ and $L_X=5$.}
 \end{centering}
 \end{figure}

As previously mentioned we have also checked the effect of the healing on the liftings {\bf lift}$_B$ and {\bf lift}$_C$ for those macroscopic variables sets which present the largest deviation from the microscopic dynamics, see Figure~\ref{fig:healing_BC}.
Healing is expected to occur whenever a sufficiently long coarse time step is chosen. The two lifting procedures react in a quite different way to the increase of the coarse time step. 
We observe that even considering very large healing time, with {\bf lift}$_B$ we still do not recover the proper dynamics, instead for {\bf lift}$_C$ we observe a very good agreement using a healing time of the order of $1/300$ of the total time. 
The reason for such result may again be founded on the fact that with {\bf lift}$_C$ the error on the cluster distribution for species $X$ are smaller than with {\bf lift}$_B$ and such small error are corrected by short healing.

\subsection{Example \ref{ex:trim}}

Let us now turn to the simpler Example~\ref{ex:trim}, in which only the nonlinear reaction is present. A direct microscopic simulation shows that a $1$-dimensional ring, initially completely covered by $X$ evolves towards a macroscopic steady state with $a_{\rm eq} \simeq 0.28$. 
If, instead, we look at the solutions of the approximating mean field equations (\ref{MF}), we obtain a steady state with equal concentrations of both species, and consequently $ a_{\rm eq} \simeq = x_{\rm eq} =0.5$.
Obviously, a more sophisticated approach is here also necessary.
	
\begin{figure}
 \begin{centering}
\includegraphics[width=0.49\textwidth]{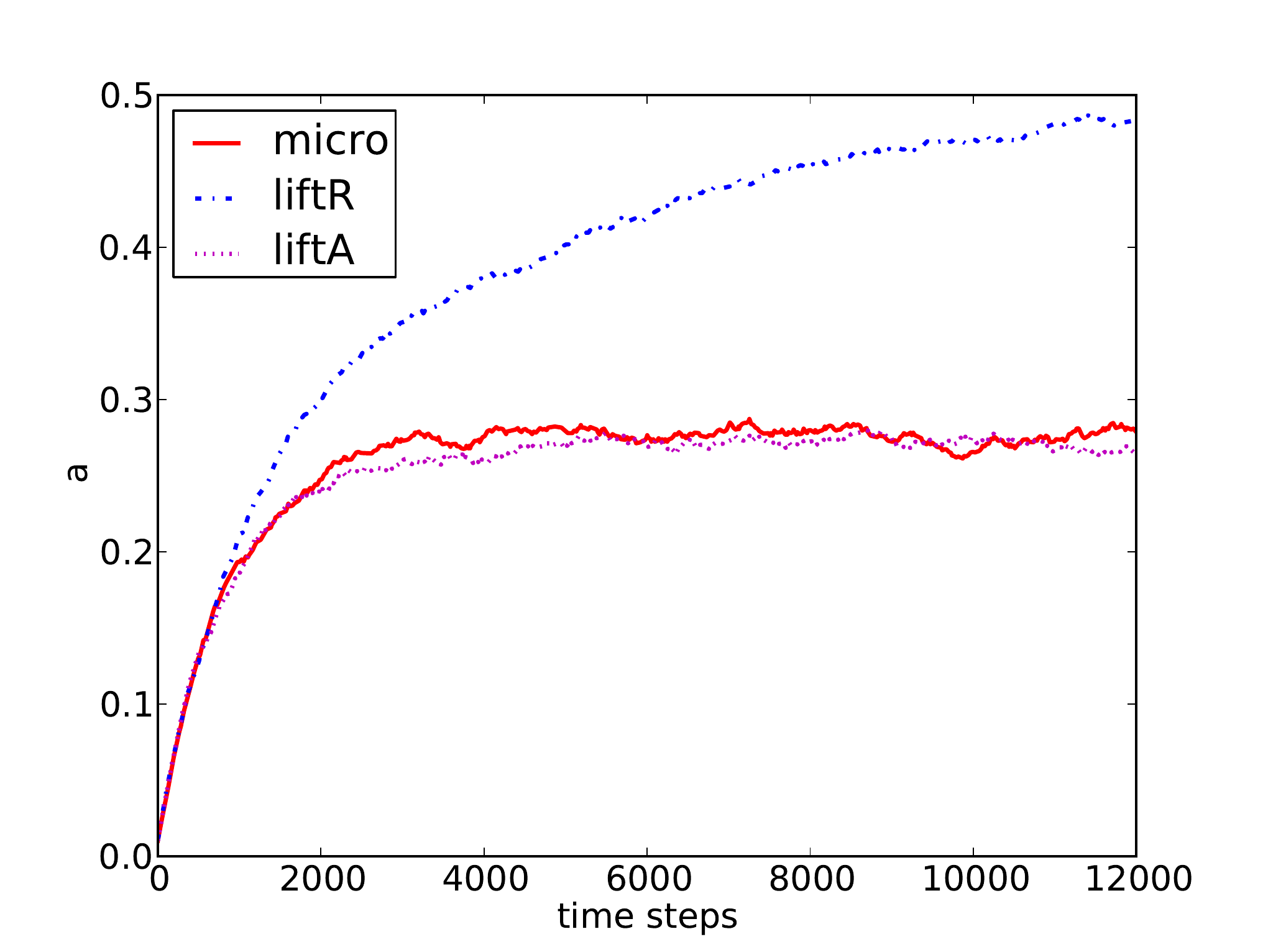}
\caption{ \label{fig:mean}Comparison of micro dynamics  with {\bf lift}$_R$ and {\bf lift}$_A$, both using $L_A=0$, $L_X=0$.}
 \end{centering}
 \end{figure}

We first consider the lifting {\bf lift$_R$}, see Remark~\ref{rem:lift_r}, which only takes into account the coverage of species $A$, and perform the same numerical experiment as for Example~\ref{ex:schl}, see Section~\ref{sec:num_schl}. The results are shown in Figure~\ref{fig:mean}.  We observe that, as expected, {\bf lift}$_R$ has $a=0.5$ as steady state, demonstrating that this lifting induces the same closure approximation as the mean field equations.

A significant improvement can be obtained by making the following observation: if we start from a uniform $X$ coverage, then any element of species $A$ appearing in the ring is always necessarily surrounded by two elements of species $X$. Therefore, any cluster longer than a isolated $A$ is physically impossible.  As a consequence, when we calculate the coverage, we are already calculating $M^1_A$, and we can easily take $\{a,M^1_A\}$ as the macroscopic variables set, instead of the sole coverage $a$.  Note that, for this choice of macroscopic state variables, all lifting procedures introduced in Section~\ref{sec:lift_proc} become very similar.  Indeed, for any value of $L_X$, the remainder of elements of species $A$ that are not accounted for by the macroscopic state variables is empty ($B^r_A=\emptyset$).  The only choice that needs to be made is the division of the elements of $B^r_X$ into clusters of different length that are necessary to separate all elements of species $A$.  Here, we choose to simply separate each two elements of species $A$ by a single cluster of species $X$ of length $L_X+1$. The remaining elements of species $X$ are then inserted randomly next to an element of species $A$. We call the resulting lifting operator {\bf lift}$_A$.

The macroscopic dynamics that results  from the coarse time-stepper is also depicted in Figure~\ref{fig:mean}. The result illustrates that it is enough to take out the physically impossible states to get an excellent agreement between the microscopic and the coarse-grained dynamics. We reiterate that both liftings essentially start from the same macroscopic information, namely the number of elements of species $A$ present in the system; the only difference is that, for \textbf{lift}$_A$, we have used some additional information about the specific features of the microscopic dynamics, namely the fact that $a\equiv M_A^1$.

\section{Discussion and outlook}\label{sec:concl}
In this paper, we have presented and studied an approach to implement numerical closure strategies,  aimed at reproducing the dynamical behavior of stochastic reactive systems taking place on low-dimensional supports (here, on a lattice).
Such closure is desired in the framework of multi-scale simulations, where one needs to accelerate the original, fully microscopic simulations. It is also particularly relevant for performing ``upper-level'' tasks on the system, such as bifurcation analysis or control design.   
Performing these tasks is often challenging in the case of low-dimensional systems, as it is not possible to use macroscopic approaches like the mean field approximation to have a reliable macroscopic description of the dynamics.  
The multi-scale strategies we used here are based on the idea of constructing a coarse time-stepper for a suitable, preferentially small set of macroscopic state variables. Microscopic simulations are then only called for to estimate the (local) time derivative of such macroscopic variables.  
The main difficulty in such an approach lies in the definition of a suitable lifting operator that maps 
the macroscopic state variables to a given microscopic configuration.
The investigation of different choices for the lifting was exactly the focus of the present paper. In particular, the main goal was to investigate which lifting operator recovers the macroscopic dynamics of the system with a minimal number of macroscopic variables.

We thus investigated several approaches using the number of clusters of particles as macroscopic variables for the system. 
This choice allows for a relatively straightforward implementation of a lifting operator, which then consists in first placing such elementary blocks on the lattice, and then filling the rest of the system with larger clusters constructed with the remaining particles. 
We proposed and analyzed several lifting procedures that differ in the way they estimate the size and the amount of such larger blocks. 
We performed several numerical experiments with these different approaches, which demonstrated how few variables might be sufficient to obtain a good agreement between the macroscopic and the microscopic dynamics.
In particular, with the same set of macroscopic variables we may obtain significantly different results using different lifting strategies. 


We specifically identified two main possible ways to perform the lifting. There exist ``simple" strategies where it is assumed that nothing  is known on the microscopic dynamics and the population of larger clusters is estimated in a ``rough'' way. 
There are also ``complex" strategies,  where more information on the microscopic behavior is taken into accountn such as a simulation-based estimate of the cluster size distribution.
For the ``simple" lifting we note that the accuracy always increases when increasing the size of the macroscopic variables set, but that at the same time a high number of them is needed to get a satisfactory accuracy.
In the case of ``complex" liftings, we observe a general better agreement with the microscopic data points, but it is possible that convergence to the microscopic dynamics is non-monotonic as a function of the number of macroscopic state variables.
Within these possible approaches we have also studied the effect of applying a long coarse time step, allowing the system to ``heal'' between different lifting procedures.
We showed how, for some liftings, relying on healing is not sufficient to correct errors coming from the reconstruction, but also that in some cases the healing substantially improve the macroscopic results.

Although we could identify efficient lifting strategies for the case hereby considered, it is not entirely clear how to best choose such strategy in the most general case. It seems that there exists a trade-off between the number of macroscopic state variables that one retains, versus the amount of effort that is put in the initialization of the remaining degrees of freedom of the system. 
An interesting possibility would be to construct an approximate distribution of large clusters, based on the most popular analytical closure schemes (such as the Kirkwood approximation, the Ursell expansion, etc.) and then rely on the healing of the system during the coarse time step to correct the (hopefully) small errors introduced in this way. This approach would in addition allow one to formalize better the action of the lifting strategies and help choosing the best procedure, based on the known physical properties of the system. 
The efficiency of a particular choice of numerical closure is not only  relevant for simulation purposes, but also give important information on what closure relations make sense in the system under consideration. 
It is our opinion that the establishment of such a connection would open the way to a systematic numerical investigation of the level of accuracy of existing possible closures in stochastic systems.
\begin{acknowledgments}
YDD and GS thank Yannis Kevrekidis for stimulating discussions. GS is a Postdoctoral Fellow of the Research Foundation-- Flanders (FWO). The work of YDD was partially funded by the Fonds National pour la Recherche Scientifique (F.N.R.S.). This work was partially supported by the Research Council of the K.U. Leuventhrough grant OT/09/27, by the Interuniversity Attraction Poles Programme of the Belgian Science Policy Office through grant IUAP/V/22.

\end{acknowledgments}
\bibliographystyle{plain}
\bibliography{myrefs}
\end{document}